\documentclass[paper,toc,nohyper]{JHEP3}
\usepackage{epsfig,cite}
\usepackage{amsbsy}
\usepackage{graphicx}
\graphicspath{{plots/}} 

\newcommand\new{\newcommand}         

\def\beq{\begin{equation}}
\def\eeq{\end{equation}}
\def\bea{\begin{eqnarray}}
\def\eea{\end{eqnarray}}

\def\d{{\rm d}}

\def\bt{B_T}
\def\bw{B_W}

\def\d{\hbox{d}}

\def\ln{\hbox{ln}}

\renewcommand{\textfraction}{0.0}

\new{\emem}{{\ifmmode\mathrm{e}^-\else e$^-$\fi}}
\new{\epem}{{\ifmmode\mathrm{e}^+\else e$^+$\fi}}
\new{\zo}  {{\ifmmode\mathrm{Z}\else Z\fi}}
\new{\epm} {{\ifmmode\mathrm{e^+e^-}\else $\mathrm{e^+e^-}$\fi}}
\new{\qq}  {{\ifmmode\mathrm{q}\else q\fi}}
\new{\qqb} {{\ifmmode\bar{\mathrm{q}}\else $\bar{\mathrm{q}}$\fi}}
\new{\bq}  {{\ifmmode\mathrm{b}\else b\fi}}
\new{\bqb} {{\ifmmode\bar{\mathrm{b}}\else $\bar{\mathrm{b}}$\fi}}
\new{\qqbar}{\qq\qqb}

\new{\LEP}        {\mbox{\small\textsc{LEP}}}
\new{\LEPONE}     {\mbox{\small\textsc{LEP1}}}
\new{\LEPTWO}     {\mbox{\small\textsc{LEP2}}}
\new{\CERN}       {\mbox{\small\textsc{CERN}}}
\new{\ALEPH}      {\mbox{\small\textsc{ALEPH}}}
\new{\DELPHI}     {\mbox{\small\textsc{DELPHI}}}
\new{\LD}         {\mbox{\small\textsc{L3}}}
\new{\OPAL}       {\mbox{\small\textsc{OPAL}}}

\new{\eV}         {{\ifmmode {\mathrm{ eV}}\else ${\mathrm{ eV}}$\fi}}
\new{\MeV}        {{\ifmmode {\mathrm{ MeV}}\else ${\mathrm{ MeV}}$\fi}}
\new{\MeVc}       {{\ifmmode {\mathrm{ MeV}}/c\else ${\mathrm{ MeV}}/c$\fi}}
\new{\MeVcc}      {{\ifmmode {\mathrm{ MeV}}/c^2\else ${\mathrm{ MeV}}/c^2$\fi}}
\new{\GeV}        {{\ifmmode {\mathrm{ GeV}}\else ${\mathrm{ GeV}}$\fi}}
\new{\GeVc}       {{\ifmmode {\mathrm{ GeV}}/c\else ${\mathrm{GeV}}/c$\fi}}
\new{\GeVcc}      {{\ifmmode {\mathrm{ GeV}}/c^2\else ${\mathrm{GeV}}/c^2$\fi}}
\new{\TeV}        {{\ifmmode {\mathrm{ TeV}}\else ${\mathrm{ TeV}}$\fi}}

\new{\JS}         {\mbox{\small\textsc{JETSET}}}
\new{\HW}         {\mbox{\small\textsc{HERWIG}}}
\new{\AR}         {\mbox{\small\textsc{ARIADNE}}}
\new{\PY}         {\mbox{\small\textsc{PYTHIA}}}
\new{\HWpp}       {\mbox{\small\textsc{HERWIG++}}}
\new{\JSv}        {\mbox{\small\textsc{JETSET\ 7.405}}}
\new{\HWo}        {\mbox{\small\textsc{HERWIG\ 5.8}}}
\new{\HWn}        {\mbox{\small\textsc{HERWIG\ 5.9}}}
\new{\ARv}        {\mbox{\small\textsc{ARIADNE\ 4.05}}}
\new{\PYv}        {\mbox{\small\textsc{PYTHIA\ 5.7}}}
\new{\HWppv}        {\mbox{\small\textsc{HERWIG++\ 2.3}}}

\new{\Mh}         {{\ifmmode M_{\mathrm{ H}}
                    \else $M_{\mathrm{H}}$\fi}}

\new{\Mz}         {{\ifmmode M_{\mathrm{Z}}
                    \else $M_{\mathrm{Z}}$\fi}}
\new{\Mzsq}       {{\ifmmode M^2_{\mathrm{ Z}}
                    \else $M^2_{\mathrm{Z}}$\fi}}

\new{\as}[1]      {{\ifmmode\alpha^{#1}_s
                    \else$\alpha^{#1}_s$\fi}}
\new{\asx}[1]      {{\ifmmode a^{#1}_s
                    \else $a^{#1}_s$\fi}}
\new{\asb}[1]     {{\ifmmode\overline{\alpha}^{#1}_s
                    \else $\overline{\alpha}^{#1}_s$\fi}}
\new{\asmz}       {{\ifmmode\alpha_s(\Mz)
                    \else $\alpha_s(\Mz)$\fi}}
\new{\lqcd}       {{\ifmmode\Lambda_{\mathrm{ QCD}}
                    \else $\Lambda_{\mathrm{ QCD}}$\fi}}
\new{\lqcdsq}     {{\ifmmode\Lambda^2_{\mathrm{ QCD}}
                    \else $\Lambda^2_{\mathrm{ QCD}}$\fi}}
\new{\llla}       {{\ifmmode\Lambda_{\mathrm{ LLA}}
                    \else $\Lambda_{\mathrm{ LLA}}$\fi}}
\new{\lmsbar}[1]  {{\ifmmode \Lambda^{(#1)}_{\overline{\mathrm{MS}}}
                    \else $\Lambda^{(#1)}_{\overline{\mathrm{MS}}}$\fi}}
\new{\lmsb}       {{\ifmmode \Lambda_{\overline{\mathrm{MS}}}
                    \else $\Lambda_{\overline{\mathrm{MS}}}$\fi}}
\new{\lmsbsq}     {{\ifmmode \Lambda^{2}_{\overline{\mathrm{MS}}}
                    \else $\Lambda^{2}_{\overline{\mathrm{MS}}}$\fi}}

\title{\boldmath  Determination of the strong coupling constant using
matched NNLO+NLLA predictions for hadronic event shapes in
$\mathrm{e}^+\mathrm{e}^-$ annihilations
}
\author{G.\ Dissertori\\
Institute for Particle Physics, ETH Zurich,\\
      8093 Zurich, Switzerland\\
    E-mail: \email{dissertori@phys.ethz.ch}}
\author{
A.~Gehrmann--De Ridder\\
Institute for Theoretical Physics, ETH Zurich,\\
      8093 Zurich, Switzerland\\
E-mail: \email{gehra@phys.ethz.ch}}
\author{
T.~Gehrmann\\
Institut f\"ur Theoretische Physik, Universit\"at Z\"urich,
Winterthurerstrasse 190,\\ CH-8057 Z\"urich, Switzerland\\
E-mail: \email{thomas.gehrmann@physik.unizh.ch}}
\author{E.W.N.~Glover\\
Institute for Particle Physics Phenomenology,
        Department of Physics,\\
        University of Durham, Durham, DH1 3LE, UK\\
    E-mail: \email{e.w.n.glover@durham.ac.uk}}
\author{
G.~Heinrich\\
Institute for Particle Physics Phenomenology,
        Department of Physics,\\
        University of Durham, Durham, DH1 3LE, UK\\
E-mail: \email{gudrun.heinrich@durham.ac.uk}}
\author{
G.~Luisoni\\
Institut f\"ur Theoretische Physik, Universit\"at Z\"urich,
Winterthurerstrasse 190,\\ CH-8057 Z\"urich, Switzerland\\
E-mail: \email{luisonig@physik.unizh.ch}}
\author{
H.~Stenzel\\
II. Physikalisches Institut, Justus-Liebig Universit\"at Giessen\\
Heinrich-Buff Ring 16, D-35392 Giessen, Germany\\
E-mail: \email{Hasko.Stenzel@exp2.physik.uni-giessen.de}}

\abstract{ We present a determination of the strong coupling
constant from a fit of QCD predictions for six event-shape
variables, calculated at next-to-next-to-leading order (NNLO) and
matched to resummation in the next-to-leading-logarithmic
approximation (NLLA). These event shapes have been measured in \epm\
annihilations at LEP, where the data we use have been collected by
the ALEPH detector at centre-of-mass energies between 91 and 206
GeV. Compared to purely fixed order  NNLO fits, we observe
that the central fit values are hardly affected, but the systematic
uncertainty is larger because the NLLA part re-introduces relatively
large uncertainties from scale variations. By combining the results
for six event-shape variables and eight centre-of-mass energies, we
find
\begin{center}
    $\asmz = 0.1224
    \;\pm\; 0.0009\,\mathrm{(stat)}
    \;\pm\; 0.0009\,\mathrm{(exp)}
    \;\pm\; 0.0012\,\mathrm{(had)}
    \;\pm\; 0.0035\,\mathrm{(theo)}$,
\end{center}
which improves previously published measurements at NLO+NLLA. We 
also carry out a detailed investigation of hadronisation corrections, using
a large set of  Monte Carlo generator predictions.}

\keywords{QCD, Jets, LEP Physics, NLO and NNLO Computations, resummation, strong coupling constant}
\preprint{{ZU-TH 09/09}, IPPP/09/42, DCPT/09/84,
{ETHZ-IPP-2009-09}}

\begin{document}


\section{Introduction}
\label{sec:intro}

Event-shape distributions in $\mathrm{e^+e^-}$ annihilation have
been measured with high accuracy at LEP at centre-of-mass energies
between 91 and 206\,GeV~\cite{ALEPH-qcdpaper,aleph,opal,l3,delphi}
and at the SLAC-SLD experiment at 91\,GeV~\cite{sld}, as well as at
the  DESY PETRA collider at lower energies, e.g. by the JADE
experiment~\cite{jade}. Event-shape observables are infrared-safe
variables designed to describe the structure of the hadronic final
state. At leading order in perturbation theory, \epm\ annihilation
to hadrons occurs via $\epm\rightarrow\qqbar$ and subsequent
hadronisation to stable hadrons, resulting in a back-to-back
(two-jet-like) structure of the event. At higher orders,  gluon
radiation off quarks will lead to deviations from this two-jet
structure.

Fixed-order QCD corrections to event-shape distributions were
calculated some time ago at next-to-leading order
(NLO)~\cite{ERT,kunszt,event}, and more recently at
next-to-next-to-leading order (NNLO)~\cite{ourt,our3j,ourevent,Weinzierl:2008iv,Weinzierl:2009ms,Weinzierl:2009nz} for
the six event-shape observables thrust $T$~\cite{farhi}
(respectively $\tau = 1-T$), heavy jet mass $M_H$~\cite{mh}, wide
and total jet broadening $B_W$ and $B_T$~\cite{bwbt},
$C$-parameter~\cite{c} and the two-to-three-jet transition parameter
in the Durham algorithm, $-\ln y_3$~\cite{durham}. The definitions
of these variables, which we denote collectively as $y$ in the
following, are summarised e.g.
in~\cite{as_theory-uncertainties,ourevent}.

As  the fixed order expansion is reliable only if the
event-shape variable is sufficiently far away from its two-jet
limit, i.e. away from $y\to 0$. This is because large logarithmic 
corrections spoil the
convergence of the perturbative expansion in the two-jet region, indicating the sensitivity to multiple soft gluon
radiation. To obtain a reliable theoretical prediction in the full
kinematical range, it is therefore necessary to resum these
logarithms to all orders in perturbation theory and to match the
resummed result to the fixed order calculation.

Until very recently, the theoretical state-of-the-art description of
event-shape distributions over the full kinematic range was based on
the matching of the next-to-leading-logarithmic approximation
(NLLA)~\cite{resumall} onto the fixed next-to-leading
order~\cite{ERT,kunszt,event} calculation. Now that the NNLO results
are available, the matching of the resummed result in the
next-to-leading-logarithmic approximation onto the NNLO calculation
has been performed~\cite{Gehrmann:2008kh} for all six event shapes
mentioned above in the so-called $\ln\,R$-matching
scheme~\cite{resumall}. It was found that the difference between
NLLA+NNLO and NNLO is largely restricted to the two-jet region,
while NLLA+NLO differed from pure NLO in normalisation throughout
the full kinematical range.
For the thrust distribution, logarithmic corrections 
reaching beyond the NLL approximation
have been calculated  recently~\cite{Becher:2008cf} using
Soft-Collinear Effective Theory (SCET)~\cite{scet}.

In the theoretical description of event-shape observables within
perturbative QCD, the only free parameter is the strong coupling
constant \as{}, such that a fit of QCD predictions to the data for
these observables lends itself to determine the strong coupling
constant with high precision. 
Several determinations of \as{} based on NNLO results have 
been performed recently: using NNLO predictions~\cite{ourevent} 
for the six event shapes listed above, a determination of \as{} 
based on ALEPH data~\cite{ALEPH-qcdpaper} has been
performed~\cite{Dissertori:2007xa}, where the systematic uncertainty
from renormalisation scale variations was found to be reduced by a
factor of two as compared to the fit based on NLO predictions
only. 
After  the NNLO+NLLA calculations for the six event shapes have 
become available~\cite{Gehrmann:2008kh},  
a determination of \as{} 
based on JADE data has been carried out in~\cite{Bethke:2008hf}.

Further, there are several studies based on thrust only:  
Using ALEPH and OPAL data for the thrust distribution and combining 
the theoretical NNLO prediction with infrared logarithms
resummed within the SCET formalism, 
a  precise determination of \as{} has also been performed 
in~\cite{Becher:2008cf}.
A re-evaluation of the non-perturbative contribution to the thrust distribution
is presented in~\cite{Davison:2008vx}, where the NNLO results of \cite{ourt,ourevent} have been matched to resummation at NLL accuracy 
and then used for a combined determination of both the low-scale effective 
coupling $\alpha_0$ and \asmz{}.

A very recent study of non-perturbative corrections based on moments of event
shapes has been carried out in \cite{Pahl:2008uc,Pahl:2009aa},
ref.~\cite{Pahl:2009aa}
containing also a determination of both $\alpha_0$ and \asmz{}.
However, these studies are based on NLO calculations only. 
NNLO predictions for event shape moments   
can be found in \cite{Ridder:2009dp}.

The agreement in the two-jet region of the matched prediction with hadron-level data
is still far from being perfect, the discrepancy being attributed mainly to
non-perturbative hadronisation corrections, but also to missing subleading
logarithms, electroweak corrections and quark mass effects.
In fact, based on the results of~\cite{Becher:2008cf}, one can estimate that the subleading 
logarithms can account for
roughly half of the discrepancy between the parton-level matched NLLA+NNLO prediction and the hadron level data.

While significant progress has been made for the perturbative calculations, the non-perturbative 
corrections for hadronisation needed to extract the value of $\alpha_s$ from event-shape distributions 
are still obtained from Monte Carlo event generators based on leading-logarithmic (LL) parton showers and 
fragmentation models \cite{ALEPH-qcdpaper,Dissertori:2007xa}. The hadronisation itself is presently parametrised by string- or cluster 
fragmentation models but the simulation of the multi-parton final state can now be performed at NLO+LL,
which is in principle more consistent with the NNLO+NLLA calculation we use in our fits. We therefore
 investigate the performance of this type of generators taking
  HERWIG++~\cite{HERWIG++} as reference, which 
represents a modern event generator allowing optionally the inclusion of NLO calculations according to different schemes.  

In this paper, we present a determination of the strong coupling
constant based on matched NLLA+NNLO results for six event-shape
variables.
Hadronisation corrections and  quark mass effects (at least to
NLO~\cite{quarkmass}) have been included in the procedure. 
We first review the theoretical framework in section \ref{sec:theory} before proceeding to the comparison with data in section \ref{sec:fit}.
The
method used for the fit follows closely that described
in~\cite{Dissertori:2007xa,ALEPH-qcdpaper}, but some improvements to
the method used in \cite{ALEPH-qcdpaper} have been made which will
be explained in section \ref{sec:fit}. Systematic
uncertainties for individual determinations of
$\alpha_s\left(Q\right)$ from different variables at different
energies are presented in section \ref{sec:syst}. Combined results
are given in section \ref{sec:results} and further systematic
studies concerning the hadronisation corrections, the scheme of matching NLLA to NNLO, the normalisation 
and quark mass 
correction procedure as well as the combination method are discussed
in section \ref{sec:studies}. Finally our findings are summarized in
section~\ref{sec:discussion}.


\section{Theoretical Framework}
\label{sec:theory}

The fixed-order QCD description of the experimentally measured
event-shape distributions
$$
\frac{1}{\sigma_{{\rm had}}}\, \frac{\d\sigma}{\d y}
$$
starts from the perturbative expansion
\begin{eqnarray}
\frac{1}{\sigma_{0}}\, \frac{\d\sigma}{\d y}
(y,Q,\mu) &=& \bar\alpha_s (\mu) \frac{\d {A}}{\d y}(y)
+ \bar\alpha_s^2 (\mu) \frac{\d {B}}{\d y} (y,x_\mu) +
\bar\alpha_s^3 (\mu) \frac{\d {C}}{\d y}(y,x_\mu) +
{\cal O}(\bar\alpha_s^4)\;, \label{eq:NNLO0mu}
\end{eqnarray}
where
$$
  \bar\alpha_s = \frac{\alpha_s}{2\pi}\;, \qquad
x_\mu = \frac{\mu}{Q}\;,
$$
and where $A$, $B$ and $C$ are the perturbatively calculated
coefficients~\cite{ourevent} at LO, NLO and NNLO.
 They have been computed with
the  parton-level event generator program {\tt EERAD3}, which
contains the relevant matrix elements with up to five external
partons~\cite{3jme,muw2,V4p,tree5p}, combined using an infrared
antenna subtraction method~\cite{ourant}. A recently discovered
inconsistency in the treatment of large-angle soft
radiation~\cite{Weinzierl:2008iv} in the original {\tt EERAD3}
implementation has been corrected, resulting in numerically minor
changes to the NNLO coefficient in the kinematical region relevant
to the phenomenological studies here. In the deep two-jet region,
e.g.\ $(1-T)\ll 0.02$, these soft correction terms turn out to be
numerically significant. They account for an initially observed
discrepancy between the {\tt EERAD3} results and the logarithmic
contributions (computed within SCET) to the thrust distribution to
 NNLO~\cite{Becher:2008cf},
which are now in full agreement.

All coefficients are normalised to the tree-level cross section for
$\epm\rightarrow\qqbar$, $\sigma_{0}$. For massless quarks, this
normalisation cancels all electroweak coupling factors, and the
dependence of (\ref{eq:NNLO0mu}) on the collision energy is only
through $\as{}$ and $x_\mu$. Summation over massless quark flavours
in $\sigma_{0}$ and $\d\sigma/\d y$ results in a constant
normalisation factor which cancels exactly in the ratio of these
quantities.

Predictions for the experimentally measured event-shape distributions
are then obtained by normalising to $\sigma_{{\rm had}}$ as
\begin{equation}
\frac{1}{\sigma_{{\rm had}}}\, \frac{\d\sigma}{\d y}(y,Q,\mu)
= \frac{\sigma_0}{\sigma_{{\rm had}}(Q,\mu)}\,
\frac{1}{\sigma_{0}}\, \frac{\d\sigma}{\d y}(y,Q,\mu)\;.
\label{eq:NNLOmu}
\end{equation}
For massless quarks, all  electroweak coupling factors cancel
in $\sigma_0/\sigma_{{\rm had}}$.

In all expressions, the scale dependence of $\alpha_s$ is determined
according to the three-loop running:
\begin{equation}
\alpha_s(\mu^2) = \frac{2\pi}{\beta_0 L}\left( 1-
\frac{\beta_1}{\beta_0^2}\, \frac{\ln L}{L} + \frac{1}{\beta_0^2 L^2}\,
\left( \frac{\beta_1^2}{\beta_0^2}\left( \ln^2 L - \ln L - 1
\right) + \frac{\beta_2}{\beta_0}  \right) \right)\;,
\label{eq:runningas}
\end{equation}
where $L= 2\, \ln(\mu/\Lambda_{\overline{{\rm MS}}}^{(N_F)})$
and $\beta_i$ are the $\overline{{\rm MS}}$-scheme coefficients listed
in~\cite{ourevent}.

The assumption of vanishing quark masses, which was used in all
 expressions for differential distributions up to here,
is only partly justified,
especially for the LEP1 data,
where bottom quark mass corrections can be relevant
at the per cent level \cite{aleph_mb}. The effect scales
with $M_{\rm b}^2/Q^2$ and decreases to $0.2$-$0.3\%$ at 200 GeV.
We take into account bottom mass effects by
retaining the massless $N_F=5$ expressions derived above and adding
 the difference between the massless and massive LO and NLO
coefficients $A$ and $B$~\cite{quarkmass},
\begin{equation}\label{quarkmasscor}
\frac{1}{\sigma_{{\rm had}}}\, \frac{\d\sigma}{\d y}(y,Q,\mu) =
\frac{1}{\sigma_{{\rm had}}}\, \left( (1-r_b(Q))\,\frac{\d\sigma}{\d
y}_{\rm massless} +r_b(Q)\frac{\d\sigma}{\d y}_{\rm massive} \right)
\, .
\end{equation}
A pole b-quark mass $M_{\rm b}$ = 4.5\,
\GeVcc\ was used and Standard Model values were taken for the
fraction $r_b(Q)$ of ${\rm b\overline{b}}$ events.
In this case, the electroweak coupling factors no
longer cancel in the ratio $\sigma_0/\sigma_{{\rm had}}$,
and the summation over quark flavours has
to be carried out explicitly.

For $\sigma_{{\rm had}}$ an NNLO calculation
(${\cal O}(\alpha_s^2)$ in QCD)~\cite{kuhnrev}
including mass corrections for the b-quark up to
${\cal O}(\alpha_s)$, and including the leading
mass terms to ${\cal O}(\alpha_s^2)$, was used to
calculate the correction $\sigma_0/\sigma_{{\rm had}}$.
A genuine ${\cal O}(\alpha_s^3)$-expression  for (\ref{eq:NNLOmu}) can be
obtained by expanding the ratio $\sigma_0/\sigma_{{\rm had}}$, as
done in~\cite{ourevent}.

Next-to-leading order electroweak corrections to event shape distributions in 
$e^+e^-$ annihilation were computed very recently~\cite{3jew}.
Using the event selection cuts and event shape definitions 
as applied in the experimental analysis~\cite{ALEPH-qcdpaper}, one observes 
substantial electroweak corrections to $\sigma_{{\rm had}}$ and 
$\d \sigma/\d y$. These corrections cancel to a very large extent in the 
ratio (\ref{eq:NNLOmu}). In the kinematical range used in the $\as{}$
determination below, the 
total electroweak corrections are 
at the level of two per cent at LEP1 and 
at most five per cent at LEP2. Genuine weak corrections from 
virtual massive gauge boson loops or fermion loops
amount to one per mille or less at both LEP1 and LEP2.  
The corrections 
are thus  much lower than initially 
anticipated from partial calculations of higher-order 
electroweak contributions~\cite{CarloniCalame:2008qn}. 
Since the experimental 
data were corrected for photon radiation effects using PYTHIA,
 it is not straightforward to 
include the electroweak corrections, and requires further study.

The resummation of large logarithmic corrections in the $y\to 0$ limit
starts from the integrated cross section:
\begin{equation}\label{Rfixed}
R\left(y,Q,\mu\right)\,\equiv\,\frac{1}{\sigma_{{\rm had}}}\int_{0}^{y}\frac{d\sigma\left(x,Q,\mu\right)}{dx}dx,
\end{equation}
which has the following fixed-order expansion:
\begin{equation}\label{Rfixedexp}
R\left(y,Q,\mu\right)\,=\,1+\,\mathcal{A}\left(y\right)
\bar{\alpha}_{s}\left(\mu\right)\,
+\,\mathcal{B}\left(y,x_\mu\right)\bar{\alpha}_{s}^{2}\left(\mu\right)\,+\,\mathcal{C}\left(y,x_\mu\right)\bar{\alpha}_{s}^{3}\left(\mu\right)+{\cal O}(\bar\alpha_s^4)\;.
\end{equation}
The fixed-order coefficients $\mathcal{A}$, $\mathcal{B}$,
$\mathcal{C}$ can be obtained by integrating the distribution
(\ref{eq:NNLO0mu}) normalised to $\sigma_{{\rm had}}$
(\ref{eq:NNLOmu}) and using $R(y_{{\rm max}},Q,\mu)=1$ to all
orders, where $y_{{\rm max}}$ is the maximum kinematically allowed
value for the event-shape variable $y$.

In the limit $y\to 0$ one observes that the perturbative
$\alpha_{s}^n$--contribution to $R(y)$ diverges like $\alpha_s^n L^{2n}$,
with $L=-\ln\, y$ ($L=-\ln\,(y/6)$ for $y=C$).
This leading logarithmic (LL) behaviour is due to multiple soft gluon
emission at higher orders, and the LL coefficients exponentiate, such that
\begin{displaymath}
\ln\,  R(y) \sim L g_1(\alpha_s L)\;,
\end{displaymath}
where $g_1(\alpha_s L)$ is a power series in its argument.

For the event-shape observables considered here, and assuming
massless quarks, leading and  next-to-leading logarithmic (NLL)
corrections can be resummed to all orders in the coupling constant,
such that
\begin{equation}\label{eq:Rresummed}
R\left(y,Q,\mu\right)\,=\ \left(1+C_{1}\bar{\alpha}_{s}
\right)\,e^{\left(L\,g_{1}\left(\alpha_{s}L\right)
+g_{2}\left(\alpha_{s}L\right)\right)}\;,
\end{equation}
where terms beyond NLL have been consistently omitted, and
$\mu=Q$ ($x_{\mu}=1$) is used.
By differentiating expression (\ref{eq:Rresummed}) with respect to
$y$, one recovers the resummed differential event-shape
distributions, which yield an accurate description for $y\to 0$.

Closed analytic forms for the  LL and NLL resummation functions
$g_1(\alpha_s L)$, $g_2(\alpha_s L)$ are available for
$\tau$~\cite{resumt}, $M_H$~\cite{resumrho}, $B_W$ and
$B_T$~\cite{resumbwbt,resumbwbtrecoil}, $C$~\cite{resumc} and
$y_3$~\cite{resumy3a}. They can be expanded as a power series, such
that
\begin{equation}
\ln R(y,Q,\mu) = \sum_{i=1}^{\infty}\sum_{n=1}^{i+1}
G_{i,i+2-n} \bar{\alpha}_s^i L^{i+2-n}\;.
\label{eq:gexpand}
\end{equation}

In order to obtain a reliable description of the event-shape
distributions over a wide range in $y$, it is mandatory to combine
fixed order and resummed predictions. However, in order to avoid the double
counting of terms common to both, the two predictions have to be
matched onto each other. A number of different matching procedures
have been proposed in the literature and for a review we refer the reader to
Ref.~\cite{as_theory-uncertainties}. The most commonly used procedure
is the so-called $\ln\, R$-matching~\cite{resumall}. In this
particular scheme, all matching coefficients can be extracted
analytically from the resummed calculation, while most other schemes
require the numerical extraction of some of the matching
coefficients from the distributions at fixed order. Since the fixed
order calculations face numerical instabilities in the region $y\to
0$, the numerical extraction of matching coefficients is prone to
large errors. Therefore we restrict ourselves to the  $\ln\,
R$-matching, studying two different variants of the latter for the present analysis.
In the $\ln\, R$-matching scheme, the NLLA+NNLO expression is
\begin{eqnarray}\label{logRmatching}
\ln\left(R\left(y,\alpha_{S}\right)\right)&=&L\,g_{1}\left(\alpha_{s}L\right)\,+\,g_{2}\left(\alpha_{s}L\right)\\
&&+\,\bar{\alpha}_{S}\left(\mathcal{A}\left(y\right)-G_{11}L-G_{12}L^{2}\right)+{}\nonumber\\
&&+\,\bar{\alpha}_{S}^{2}\left(\mathcal{B}\left(y\right)-\frac{1}{2}\mathcal{A}^{2}\left(y\right)-G_{22}L^{2}-G_{23}L^{3}\right){}\nonumber\\
&&+\,\bar{\alpha}_{S}^{3}\left(\mathcal{C}\left(y\right)-\mathcal{A}\left(y\right)\mathcal{B}\left(y\right)+\frac{1}{3}\mathcal{A}^{3}\left(y\right)-G_{33}L^{3}-G_{34}L^{4}\right)+{\cal O}(\bar\alpha_s^4)\nonumber\;.
\end{eqnarray}
The matching coefficients appearing in this expression can
be obtained from (\ref{eq:gexpand}) and are given in ~\cite{Gehrmann:2008kh},
where we refer the reader to for more details. Quark mass corrections to this
expression are included by retaining the mass dependence of the
fixed-order coefficients ${\cal A}$ and  ${\cal B}$, which follow from
the mass-corrected coefficient functions (\ref{quarkmasscor}) and
the mass-dependent total hadronic cross section.

The coefficients in
(\ref{logRmatching}) explicitly  depend on $x_\mu$, thereby stabilising the
scale dependence of the theoretical prediction:
\begin{eqnarray}
\alpha_s & \to & \alpha_s(\mu)\;, \\
\nonumber \\
\mathcal{B}\left(y\right) &\to &
\mathcal{B}\left(y,\mu\right)=\beta_{0}\,
\ln x_\mu \, \mathcal{A}\left(y\right)
+\mathcal{B}\left(y\right)\;,\nonumber \\
\mathcal{C}\left(y\right) & \to &
\mathcal{C}\left(y,\mu\right)=\left(\beta_{0}\, \ln x_\mu
\right)^{2}\mathcal{A}\left(y\right) +\ln x_\mu
\,\left[2\,\beta_{0}
\mathcal{B}\left(y\right)+\beta_{1}\,\mathcal{A}\left(y\right)\right]
+\mathcal{C}\left(y\right)\;,\nonumber\\
\label{fixedorderrenscaledependence}\\
\nonumber\\
g_2\left(\alpha_{S}L\right) &\to &
{g}_{2}\left(\alpha_{S}L,\mu^{2}\right)
=g_{2}\left(\alpha_{S}L\right)+\frac{\beta_{0}}{2\pi}
\left(\alpha_{S}L\right)^{2}\,
g_{1}'\left(\alpha_{S}L\right)\,\ln x_\mu \;,
\label{g2mudep} \\
\nonumber \\
G_{22}&\to & G_{22}\left(\mu\right)=G_{22}\,+\,\beta_{0}G_{12}\ln x_\mu
\;,\nonumber\\
G_{33}&\to & {G}_{33}\left(\mu\right)=G_{33}\,
+\,2 \beta_{0}  G_{23}\ln x_\mu\,,
\label{Gijdeponrenorm}
\end{eqnarray}
where $g_1'$ denotes the derivative of $g_1$ with respect to its
argument. This scale variation also exemplifies a tension between
NLLA and NNLO, since the NNLO coefficients compensate the
renormalisation scale variation of $\alpha_s$ up to two loops, while
the NLLA coefficients only compensate the one-loop variation. A
fully consistent matching, including the full scale dependence, is
therefore only accomplished by combining NLLA+NLO or NNLLA+NNLO. In
order to assess the effect of this incomplete compensation of
scale-dependent terms, we have computed the two-loop terms
proportional to $x_\mu$ in the above resummation and matching
functions, i.e.\ the scale-dependent logarithms appearing in $g_3$
and in the associated matching coefficients $G_{21}$ and $G_{32}$,
and recomputed the theoretical error in this new matching scheme,
which we call the $\ln R(\mu)$-scheme. In this scheme the NLLA+NNLO
expression becomes
\begin{eqnarray}\label{logRmumatching}
\ln\left(R\left(y,\alpha_{S}\right)\right)&=&L\,g_{1}\left(\alpha_{s}L\right)\,+\,g_{2}\left(\alpha_{s}L\right)\,+\,\bar{\alpha}_{S}\,g_{3}\left(\alpha_{S}L\right)\nonumber\\
&&+\,\bar{\alpha}_{S}\left(\mathcal{A}\left(y\right)-G_{11}L-G_{12}L^{2}\right)+{}\nonumber\\
&&+\,\bar{\alpha}_{S}^{2}\left(\mathcal{B}\left(y\right)-\frac{1}{2}\mathcal{A}^{2}\left(y\right)-G_{21}L-G_{22}L^{2}-G_{23}L^{3}\right){}\nonumber\\
&&+\,\bar{\alpha}_{S}^{3}\left(\mathcal{C}\left(y\right)-\mathcal{A}\left(y\right)\mathcal{B}\left(y\right)+\frac{1}{3}\mathcal{A}^{3}\left(y\right)-G_{32}L^{2}-G_{33}L^{3}-G_{34}L^{4}\right)+{\cal O}(\bar\alpha_s^4)\,,\nonumber\\
\end{eqnarray}
where the $\mu$-dependence of $g_3$ is
\begin{eqnarray}\label{g3deponrenorm}
g_{3}\left(\alpha_{S}L\right) &\to&
g_{3}\left(\alpha_{S}L,\mu^{2}\right)=g_{3}\left(\alpha_{S}L\right)+
\left(\alpha_{S}L\right)\ln\,x_{\mu}\left[\beta_{0}g_{2}'\left(\alpha_{S}L\right)\,+
\,\frac{\beta_{1}}{2\pi}\left(\alpha_{S}L\right)g_{1}'\left(\alpha_{S}L\right)\right]\nonumber\\
&&+\frac{\beta_{0}^{2}}{\pi}\left(\alpha_{S}L\right)^{2}\left(\ln\,x_{\mu}\right)^{2}\,
\frac{d}{d\left(\alpha_{S}L\right)}\left(\frac{d}{dL}\left(L\,g_{1}\left(\alpha_{S}L\right)\right)\right)\,.
\end{eqnarray}
The "bare" $g_{3}\left(\alpha_{S}L\right)$ is not known and is put
to zero, whereas the renormalisation scale dependence is
proportional to the derivatives of $g_1$ and $g_2$. In order to have
a correct compensation of the renormalisation scale dependence the
following further substitutions are made
\begin{eqnarray}
G_{21} &\to& G_{21}\left(\mu^{2}\right)=G_{21}\,+\,\beta_{0}G_{11}\ln\,x_{\mu}\,,\nonumber\\
G_{32} &\to&
G_{32}\left(\mu^{2}\right)=G_{32}\,+\,\left(\beta_{0}\ln\,x_{\mu}\right)^{2}G_{12}\,+\,\ln\,x_{\mu}\left(2\beta_{0}G_{22}\,+\,2\beta_{1}G_{12}\right)\,.\label{G21G32deponrenorm}
\end{eqnarray}

As discussed in detail in Section~\ref{sec:studies} below,
the resulting theoretical error is considerably lower than
in the standard $\ln\, R$-matching scheme, and comparable
to the error obtained in~\cite{Becher:2008cf}, where the
thrust distribution was computed beyond NLLA and matched to
NNLO~\cite{ourt,ourevent}.

In order to ensure the vanishing of the matched expression
at the kinematical boundary $y_{\textrm{\tiny{max}}}$, the further
substitution~\cite{as_theory-uncertainties} is made:
\begin{equation}\label{Ltilde}
L\,\longrightarrow\,\tilde{L}\,
=\,\frac{1}{p}\,\ln\left(\left(\frac{y_{0}}{x_{L}\,y}\right)^{p}
-\left(\frac{y_{0}}{x_{L}\,y_{\textrm{\tiny{max}}}}\right)^{p}+1\right),
\end{equation}
where $y_0 = 6$ for $y=C$ and $y_0=1$ otherwise. The values $p=1$ and
$x_{L}=1$ are taken as default.
The arbitrariness in the choice of the logarithm to be
resummed can be quantified by varying the constant $x_{L}$.


\section{Determination of the strong coupling constant}
\label{sec:fit}

As in the analysis of Ref.~\cite{Dissertori:2007xa} we have studied
the six event-shape distributions thrust $T$ \cite{farhi}, heavy jet
mass $M_H$ \cite{mh}, total and wide jet broadening ($\bt, \bw$)
\cite{bwbt}, C-parameter $C$ \cite{c} and the two-to-three-jet
transition parameter in the Durham algorithm $-\ln y_3$
\cite{durham}. The definitions of these variables and a discussion
of their properties can be found in Refs.\ \cite{ALEPH-qcdpaper} and
\cite{as_theory-uncertainties,NNLAreview}.

The measurements have been carried out by the \ALEPH\ collaboration \cite{ALEPH-qcdpaper}
\footnote{The tables with numbers and uncertainties for all variables can be found at
{\tt http://aleph.web.cern.ch/aleph/QCD/alephqcd.html}.},
at centre-of-mass energies of 91.2, 133, 161, 172, 183, 189, 200 and
206 \GeV. Earlier measurements and complementary data sets from the
LEP experiments and from SLD can be found in
Refs.~\cite{aleph,opal,l3,delphi,sld}. The event-shape
distri\-butions were computed using the reconstructed momenta and
energies of charged and neutral particles. The measurements have
been corrected for detector effects, i.e. the final distributions
correspond to the so-called particle (or hadron) level. The particle
level is defined by stable hadrons with a lifetime longer than
$10^{-9}$ s after hadronisation and leptons according to the
definition given in \cite{aleph_mega}. In addition, at LEP2 energies
above the Z peak they were corrected for initial-state radiation
effects. At energies above 133 GeV,  backgrounds from 4-fermion
processes, mainly from W-pair production and also ZZ and
Z$\gamma^*$, were subtracted following the procedure given
in~\cite{ALEPH-qcdpaper}. The experimental uncertainties were
estimated by varying event and particle selection cuts. They are
below 1\% at LEP1 and slightly larger at LEP2. For further details
we refer to Ref.\ \cite{ALEPH-qcdpaper}.

The determination of the coupling constant from these data follows
very closely the approach chosen in Refs.~\cite{Dissertori:2007xa,
ALEPH-qcdpaper}. The perturbative predictions for the distributions,
as described in section \ref{sec:theory}, are calculated to the same
order of perturbation theory for all of these variables and fit to
the data. The measurements from several variables are combined,
since this yields a better estimator of \as{}\ than using a single
variable. Furthermore, the spread of values of \as{}\ is an
independent estimate of the theoretical uncertainty. At
centre-of-mass energies above the Z peak the statistical
uncertainties are larger and background conditions are more
difficult than at the peak. Therefore a combination of
measurements is particularly important for those energies. We apply
the same combination procedure as described in
\cite{Dissertori:2007xa, ALEPH-qcdpaper}, which is based on weighted
averages and takes into account correlations between the event-shape
variables. However, in this paper we also investigate a combination
procedure which excludes the perturbative uncertainty as weight, in
order to evaluate the stability of our nominal combination method.
Note that for energies above \Mz\ we adopt the same treatment of
statistical uncertainties as in \cite{Dissertori:2007xa,
ALEPH-qcdpaper}. The same holds for the fit ranges chosen at the
LEP2 energies \cite{Dissertori:2007xa}, whereas for the LEP1 data
the fit ranges have been slightly extended, motivated by the
expected better description of the two-jet range by the resummed
calculations.

In this paper we present  fits of matched
NNLO+NLLA predictions and compare them to pure NNLO and matched
NLO+NLLA calculations as used in the analyses of Refs.\
\cite{Dissertori:2007xa, ALEPH-qcdpaper}. The nominal value for the
renormalisation scale $x_\mu = \mu/Q$ is unity. The perturbative QCD
prediction is corrected for hadronisation and resonance decays  by
means of a transition matrix, which is computed with the Monte Carlo
generators PYTHIA \cite{Pythia}, HERWIG \cite{Herwig}  and ARIADNE
\cite{Ariadne}, all tuned to global hadronic observables at $\Mz$
\cite{aleph_mega}. The parton level is defined by the quarks and
gluons present at the end of the parton shower in PYTHIA and HERWIG
and the partons resulting from the colour dipole radiation in
ARIADNE. Corrected measurements of event-shape distributions are
compared to the theoretical calculation at particle level.
In section \ref{sec:studies} we investigate the use of the NLO+LL event generator HERWIG++.

The value of \as{}\ is determined at each energy using a binned
least-squares fit. The fit programs of  Ref.\ \cite{Dissertori:2007xa}
have been extended to incorporate the NNLO+NLLA calculations.
Only statistical uncertainties arising from
the limited number of observed events, from the number of simulated
events used to calculate hadronisation and detector corrections and
from the integration procedure of the NNLO coefficient functions
are included in the $\chi^2$ of the fit. Its quality is good for all variables
at all energies. Nominal results for thrust and $-\ln y_3$, based on (\ref{logRmatching}) and using the fitted values of
\as{}, are shown in
Fig.~\ref{fig:alphas_fits1}, together
with the measured distributions. The resulting measurements of
$\as{}(Q)$ for all six event shapes are given in Table~\ref{tab:indiv1} for 91.2 to 172
GeV and in Table~\ref{tab:indiv2} for 183 to 206 GeV.
Comparisons of fits using different perturbative approximations are shown
in Fig.~\ref{fig:comp_nlo_nnlo1} for the variables thrust and $y_3$ and results for 
all variables at LEP1 are given in Table~\ref{tab:comp_nlo_nnlo}.

\FIGURE{
\begin{tabular}{lr}
\hspace*{-0.5cm}\includegraphics[width=7.7cm]{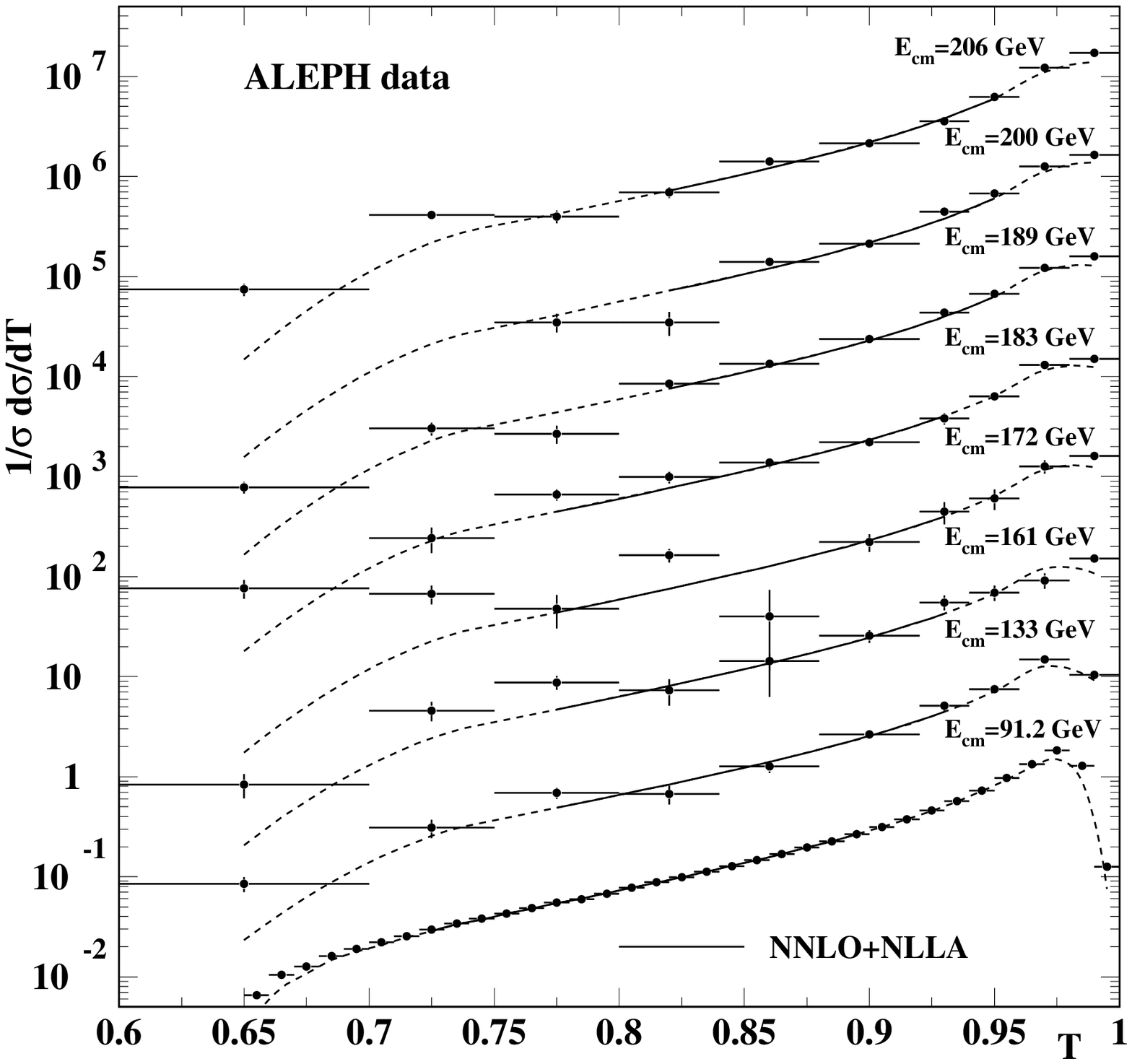} &
\hspace*{-0.5cm}\includegraphics[width=7.7cm]{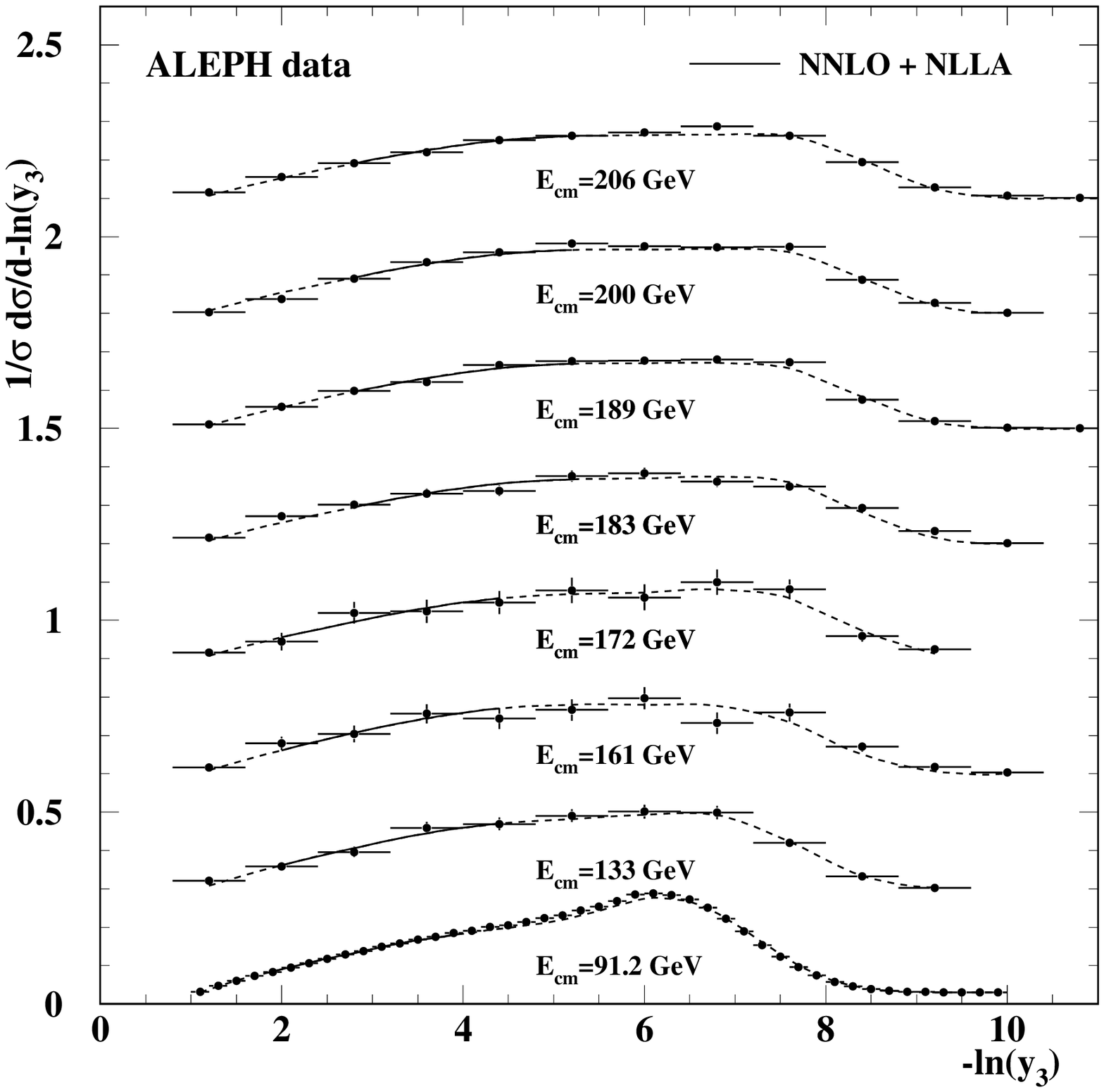}
\end{tabular}
\caption{\small Distributions measured by \ALEPH, after correction
for backgrounds and detector effects, of thrust and the
two-to-three-jet transition parameter in the Durham algorithm at
energies between 91.2 and 206 GeV, together with the fitted
NNLO+NLLA QCD predictions. The error bars correspond to the
statistical uncertainties. The fit ranges cover the central regions
indicated by the solid curves, the theoretical predictions
extrapolate well outside these fit ranges, as shown by the dotted
curves. The plotted distributions are scaled by arbitrary factors
for presentation.} \protect\label{fig:alphas_fits1} }

\FIGURE{
\begin{tabular}{lr}
\hspace*{-0.5cm}\includegraphics[width=7.7cm]{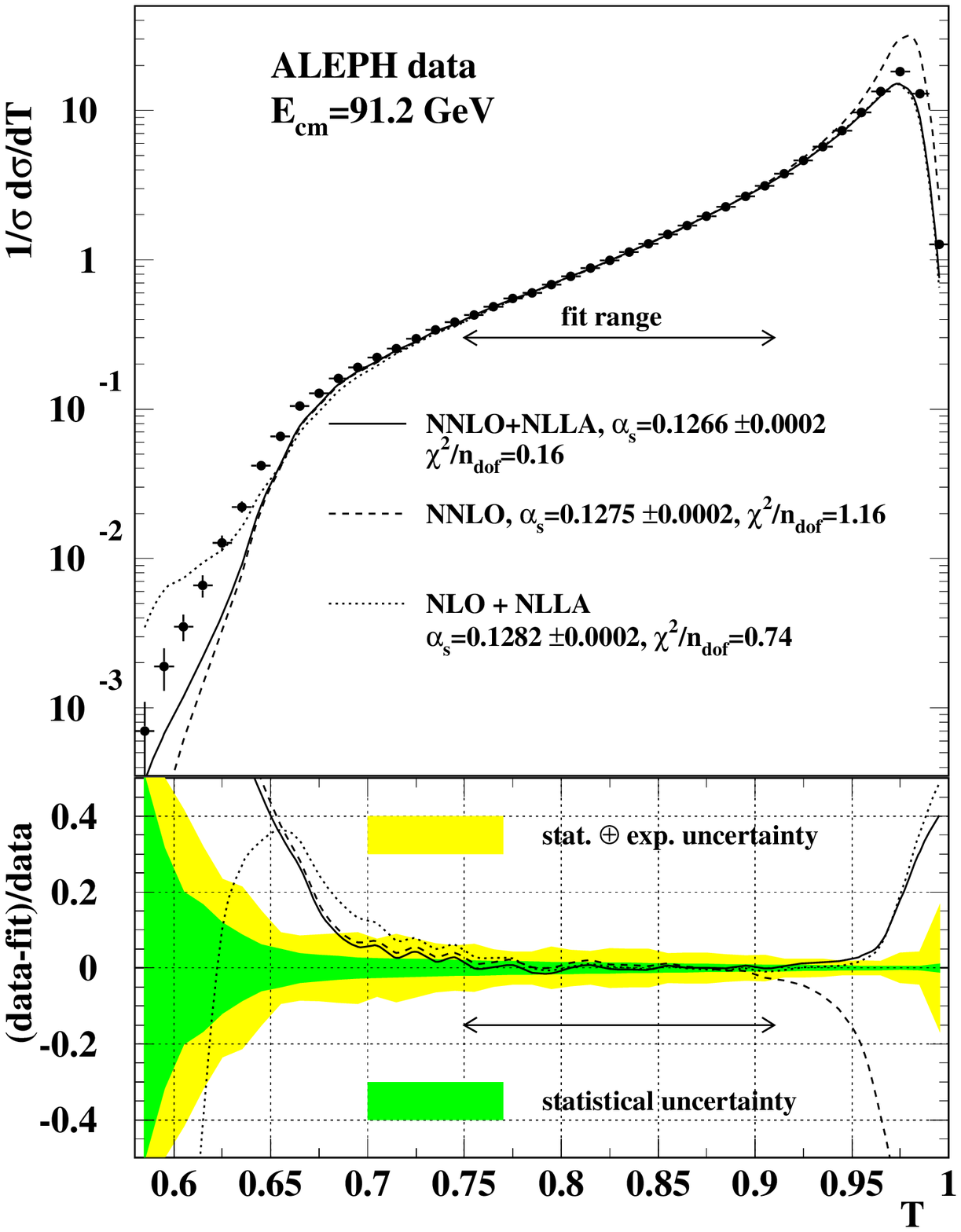} &
\hspace*{-0.5cm}\includegraphics[width=7.7cm]{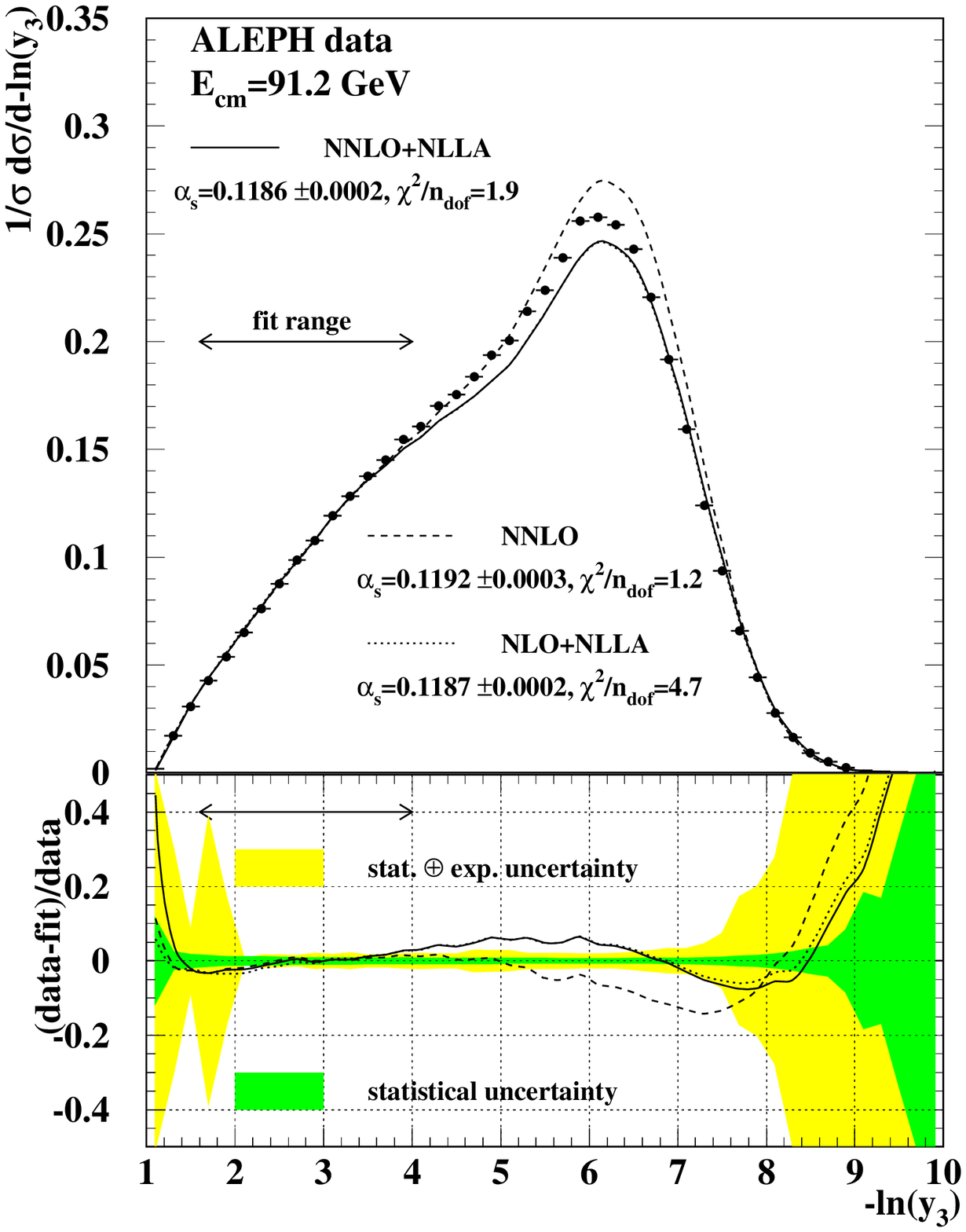}
\end{tabular}
\caption{Distributions measured by \ALEPH\ at LEP1, after correction
for detector effects, of thrust and the
two-to-three-jet transition parameter in the Durham algorithm.
Fitted QCD predictions at different orders of perturbation theory
are overlaid. The lower insets show a relative comparison of data
and QCD fits. } \protect\label{fig:comp_nlo_nnlo1} }


\section{Systematic Uncertainties of \boldmath $\alpha_s$ \unboldmath}
\label{sec:syst}


For a description of the determination and treatment of experimental
systematic uncertainty we refer to Refs.~\cite{Dissertori:2007xa,
ALEPH-qcdpaper}, since the identical approach is taken for this
analysis. Similarly, the analysis of theoretical uncertainties goes
along the lines of these earlier publications. The main source of
arbitrariness in the predictions is the choice of the
renormalisation scale $x_\mu$ and of the logarithmic rescaling
variable $x_L$. The residual dependence of the fitted value of \asmz\,\, 
on the renormalisation scale is shown in Fig.~\ref{fig:scale-var1},
for the same two variables as in the previous figures. Most notably,
the matching of NLLA terms to the NNLO prediction does not lead to a
reduced scale dependence, compared to pure NNLO only, but at least
to an improvement compared to NLO+NLLA. This could be anticipated by
the discussion in section \ref{sec:theory} on the scale dependence
of the NNLO and NLLA predictions. A further study of this particular
aspect is described in section \ref{sec:studies} below.

\FIGURE{
\begin{tabular}{lr}
\hspace*{-0.5cm}\includegraphics[width=7.7cm]{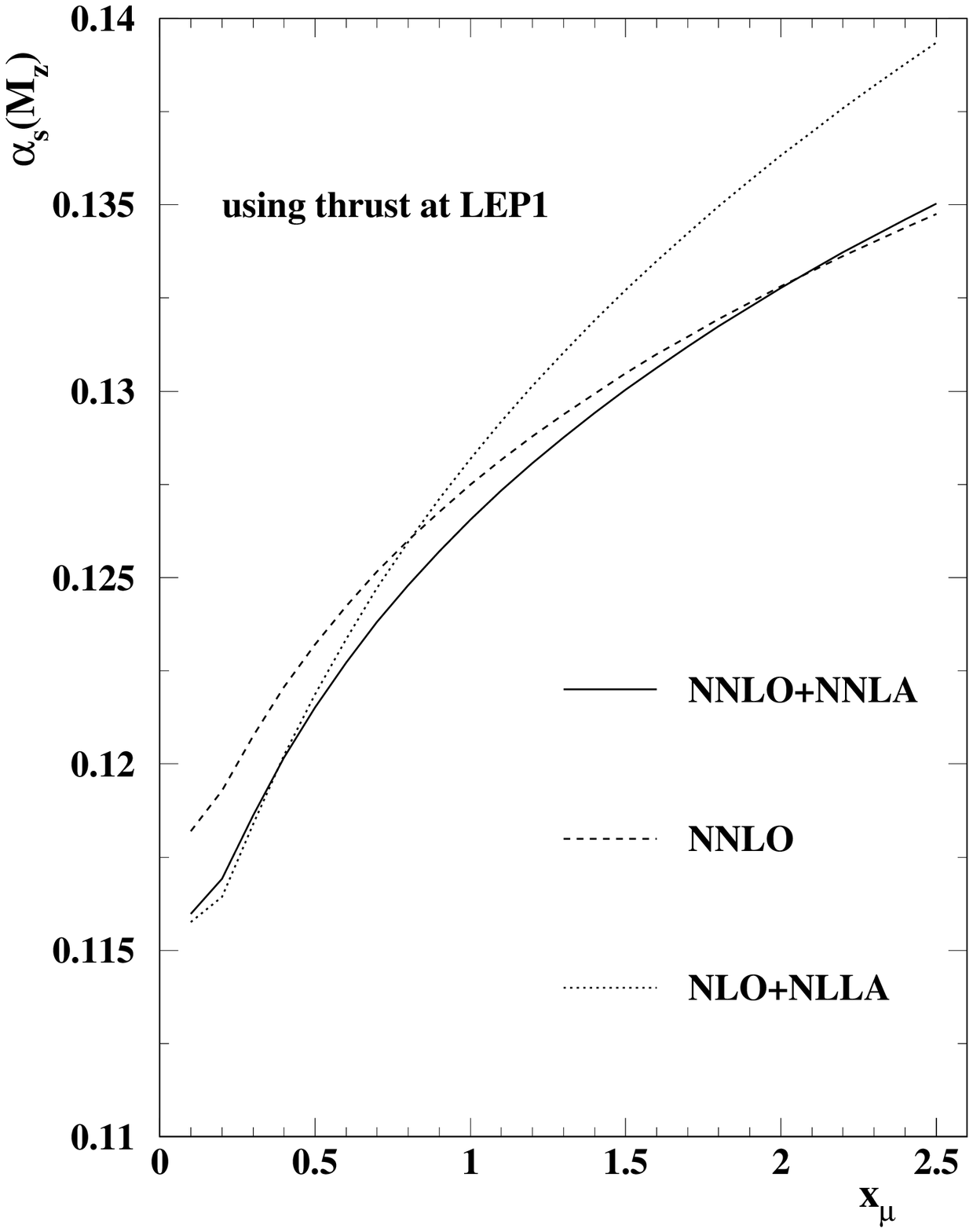} &
\hspace*{-0.5cm}\includegraphics[width=7.7cm]{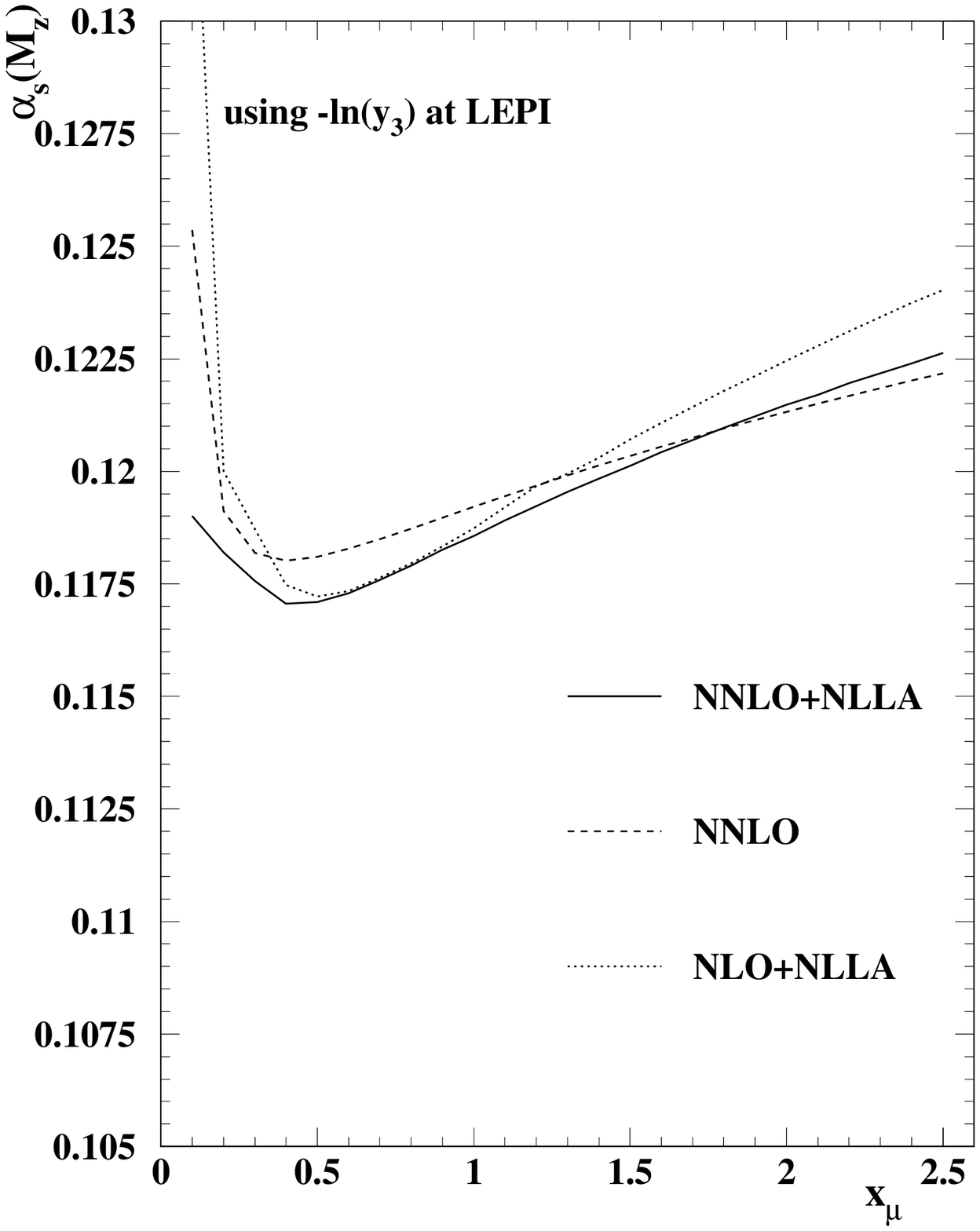}
\end{tabular}
\caption{\small Dependence of the extracted \as{}-value on the
renormalisation scale when fitting the distributions of thrust
(left) and the two-to-three-jet transition parameter in the Durham
algorithm (right) with predictions at NNLO+NLLA (solid), 
NNLO (dashed) and NLO+NLLA (dotted). } \protect\label{fig:scale-var1} }

The systematic uncertainty related to missing higher orders is
estimated with the uncertainty-band method recommended in
Ref.~\cite{as_theory-uncertainties}. Briefly, this method derives
the uncertainty of \as{}\ from the uncertainty of the theoretical
prediction for the event-shape distribution and proceeds in three
steps. First a reference perturbative prediction, here NNLO+NLLA
with $x_\mu = 1$ and $x_L=1$, is determined using the value of
\as{}\ obtained from the combination of the six variables and eight
energies, as explained in section~\ref{sec:results}. Then variants
of the prediction with different choices for $x_\mu$ and $x_L$, for
the kinematic constraint $y_{\rm max}$ and the modification degree
power $p$ are calculated with the same value of
 \as{}. A variation of the matching scheme as advocated in Ref.~\cite{as_theory-uncertainties}
was not included in the list of variants, since no $R$-matching
scheme is presently available at NNLO+NLLA. In each bin of the
distribution for a given variable, the largest upward and downward
differences with respect to the reference prediction are taken to
define an uncertainty band around the reference theory. In the last
step, the value of \as{}\ in the reference prediction is varied, in
order to find the range of values which result in predictions lying
inside the uncertainty band for the fit range under consideration.
In contrast to the original method \cite{as_theory-uncertainties} we
do not require the reference prediction to lie strictly inside the
uncertainty band, since for the present NNLO+NLLA calculations the
latter  is still subject to statistical fluctuations. Instead, we
make a fit of the reference theory with \as{} as free parameter to
the uncertainty band, which includes the statistical uncertainty on
the $C$ coefficient, as in Ref.~\cite{Dissertori:2007xa}. The values
of \as{}\ fitted to the upper and lower contour of the uncertainty
band finally set the perturbative systematic uncertainty. The upward
and downward uncertainties are very similar in magnitude and the
larger is quoted as symmetric uncertainty. The method is illustrated
in Fig. \ref{fig:band1} for thrust and the two-to-three-jet
transition parameter in the Durham algorithm $-\ln y_3$.

\FIGURE{
\begin{tabular}{lr}
\hspace*{-0.5cm}\includegraphics[width=7.7cm]{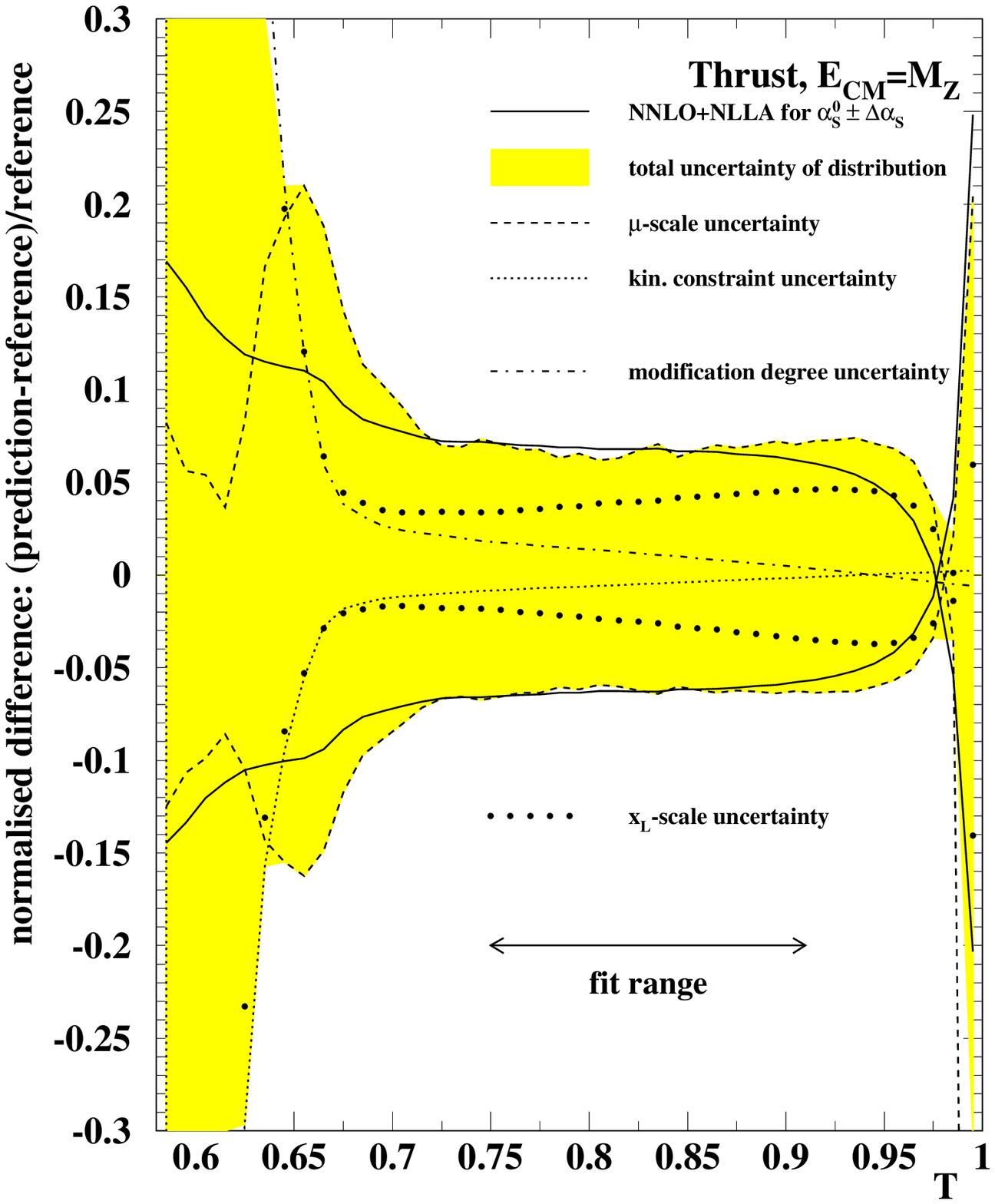} &
\hspace*{-0.5cm}\includegraphics[width=7.7cm]{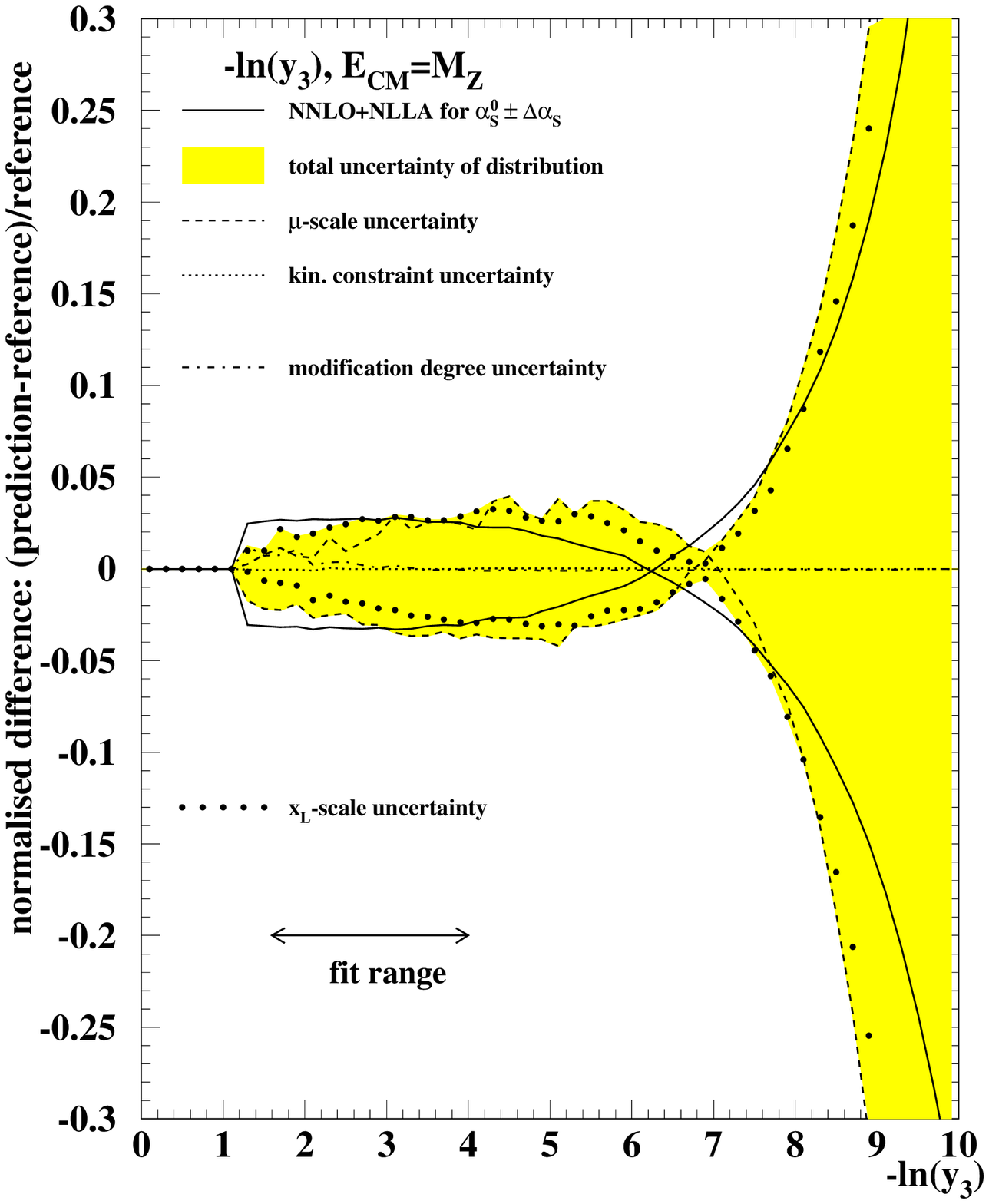}
\end{tabular}
\caption{\small  Theoretical uncertainties for the distributions of
thrust (left) and the two-to-three-jet transition parameter in the
Durham algorithm (right) at LEP1. The filled area represents the
perturbative uncertainties of the distribution for a given value
\as{0}. The curves show the reference prediction with $\alpha_s^0
\pm \Delta\alpha_s$. The theoretical uncertainty $\Delta\alpha_s$ is
derived from a fit of the reference theory to the contour of the
uncertainty band for the actual fit range. }
\protect\label{fig:band1} }

The combined value of \as{}, used to derive the systematic
perturbative error, depends itself on the theoretical error.
Hence the procedure of calculating the \as{} combination and its
perturbative error is iterated until convergence is reached,
typically after two iterations.

At LEP2 energies the statistical fluctuations are large. In order to
avoid biases from downward fluctuations, the theoretical
uncertainties are calculated with the value of \as{}\
obtained by the global combination procedure.
For each energy point and in each variable, the combined \as{}
is evolved to the appropriate energy scale and the uncertainty is
calculated for the fit range used for the different variables.

An additional error is evaluated for the b-quark mass correction
procedure. This correction has only been calculated to
${\cal O}(\as{2})$ for the differential coefficients; no resummed
and NNLO expressions are yet available. We have updated the calculations for 
the massive coefficients used in \cite{ALEPH-qcdpaper,Dissertori:2007xa} to include 
now three different sets for $M_{\rm b}$ = 4.0,\, 4.5 and 5.0  \GeVcc\,.
The difference  in \as{}\ obtained with these different sets is taken 
as systematic error. 
The difference between the massless and massive
expression for the hadronic cross section is already rather small and not
included in this estimate.
 
The total perturbative uncertainty quoted
in the tables is the quadratic sum of the errors for missing higher
orders and for the mass correction procedure. The total
perturbative error is between 3\% and 5\% at \Mz\  and decreases to
between 2\% and 3\% at LEP2 energies.

The hadronisation model uncertainty is estimated by comparing
the standard hadron-level event generator programs \HW\
and \AR\ to \PY\ for both hadronisation and detector
corrections. The same set of corrections as in
Ref.~\cite{ALEPH-qcdpaper} is used.
Both corrections are calculated with the same generator in order
to obtain a coherent description at the hadron level.
The maximum change with respect to the nominal result
using \PY\ is taken as the systematic error. At LEP2 energies the hadronisation
model uncertainty is again subject to statistical fluctuations.
These fluctuations are observed from one energy to the next and
originate from limited statistics of the fully simulated detector-correction functions. Since
non-perturbative effects are expected to decrease with $1/Q$, the
energy evolution of hadronisation errors has been fitted to a simple
$A + B/Q$ parametrisation. The fit was performed for each variable
separately. In the fit procedure a weight scaling with luminosity
is assigned to the hadronisation uncertainty at each energy point.
This ensures that the hadronisation uncertainty at \Mz,
which is basically free of statistical fluctuations, is not
altered by the procedure. As in the case of experimental systematic
uncertainties, the hadronisation uncertainty is essentially
identical to that published in \cite{ALEPH-qcdpaper}.
In section~\ref{sec:studies} we present an attempt to use modern event generators to estimate the hadronisation corrections.

The perturbative component of the
error, which is the dominant source of uncertainty in most cases, is highly correlated between the energy
points. The perturbative errors decrease with increasing $Q$, and faster than
the coupling constant itself. The overall error is in general dominated by the
renormalisation scale dependence.


\section{Combined Results}
\label{sec:results}

The measurements obtained from the six different variables using NNLO+NLLA calculations are combined into a single
measurement per energy using weighted averages.
The same procedure as in Refs.\  \cite{Dissertori:2007xa, ALEPH-qcdpaper}
is applied here. However, we investigate also
the impact of the theoretical uncertainties in the calculations of the weights, as described in section
\ref{sec:studies} below.

In Table  \ref{tab:comblep} the weighted averages are given for all LEP1 and LEP2 data, as well as for the
LEP2 data only. Essentially identical results with very similar errors are found. The fitted values of \as{}\ at the various centre-of mass energies are displayed in Fig.\ \ref{fig:comb_run} and compared to the QCD three-loop formula for the running of the coupling constant. Excellent agreement of the data with the expected
energy dependence is observed.

\FIGURE{
\hspace{1cm}
\includegraphics[width=8cm]{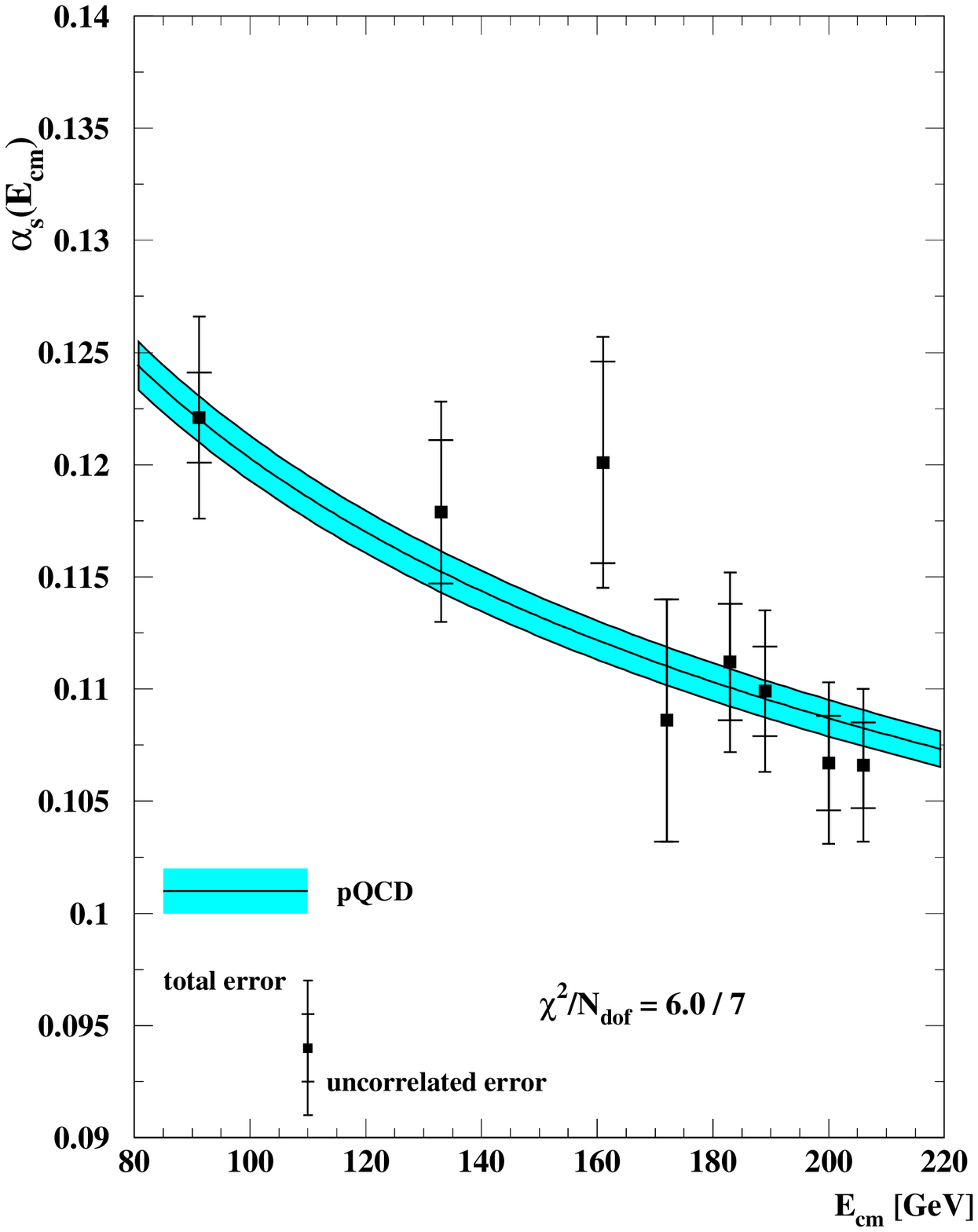}
\hspace{1cm}
\caption{\small The measurements of the strong coupling constant
$\alpha_s$ between 91.2 and 206 GeV. The results using the six
different event-shape variables are combined with correlations taken
into account. The inner error bars exclude the perturbative
uncertainty, which is expected to be highly correlated between the
measurements. The outer error bars indicate the total error. A fit
of the three-loop evolution formula using the uncorrelated errors is
shown. The shaded area corresponds to the uncertainty in the
fit parameter $\Lambda^{(5)}_{\overline {\rm MS}}=284 \pm 14 $ MeV
of the three-loop formula, eq.\
(\protect{\ref{eq:runningas}}).}\protect\label{fig:comb_run} }

In Table \ref{tab:compall} we compare the combined results obtained
at NNLO+NLLA accuracy to the results at NNLO and NLO+NLLA. The 
 numbers given in Table~\ref{tab:compall} supersede 
 those published in \cite{Dissertori:2007xa} at NNLO and NLO+NLLA and
 can be traced back
to the following changes in the present analysis
\begin{itemize}
\item for the normalisation to $\sigma_{\rm had}$ an expansion
of the ratio $\sigma_0/\sigma_{\rm had}$ was applied in
Refs.~\cite{Dissertori:2007xa,ALEPH-qcdpaper}, while here the exact
value is used;
\item new massive coefficients using $M_{\rm b}$ =\, 4.5 \GeVcc\ up to NLO are used; 
\item a massive expression for $\sigma_{\rm had}$ is now adopted;
\item for a given observable and energy, the same fit ranges (given in Tables~\ref{tab:indiv1} and \ref{tab:indiv2}) are applied to different theoretical predictions; 
\item a small transcription error in the fit program used in \cite{Dissertori:2007xa} when calculating the NNLO term for $-\ln y_3$ is corrected;
\item the previously incomplete
treatment of large-angle soft radiation~\cite{Weinzierl:2008iv}
 in {\tt EERAD3} is corrected,
resulting in minor numerical shifts in the NNLO coefficients;
\item while in \cite{Dissertori:2007xa,ALEPH-qcdpaper} the NLO+NLLA predictions were obtained by a numerical
derivative of $R$ (cf. eq. \,(\ref{Rfixed})), we now compute the
differential distributions analytically for the resummed part, which
yields a better numerical stability. We apply this procedure also
to NNLO+NLLA.
\end{itemize}

As was anticipated in section \ref{sec:theory}, the matching of the NNLO prediction with the
resummation at NLLA introduces a renormalisation scale dependence
which is absent in the pure NNLO case, 
 as described in detail in section \ref{sec:theory}. 
This is reflected by the increased perturbative and finally
total uncertainty of the NNLO+NLLA result compared to NNLO, as can be seen 
by comparing Table \ref{tab:combz} for the combined value of \asmz\ at different energies at NNLO+NLLA 
with Table \ref{tab:combz_nnlo} at NNLO.
However, compared to the NLO+NLLA fit, an improvement of more than 20\% is obtained for the
perturbative error. The central values of the fits for the different approximations turn out to be pretty
similar.
The fitted values of the coupling constant as found from  the various event-shape variables, combined
over all energies, are shown in Fig.\ \ref{fig:scatter}. Besides the larger uncertainties,
at NNLO+NLLA we observe the same reduced scatter of the results compared to NLO+NLLA as
already reported previously \cite{Dissertori:2007xa}. However, the effect is not as strong as going from a NLO fit (where the scatter is largest)
to a pure NNLO fit.

\FIGURE{
\includegraphics[width=14cm]{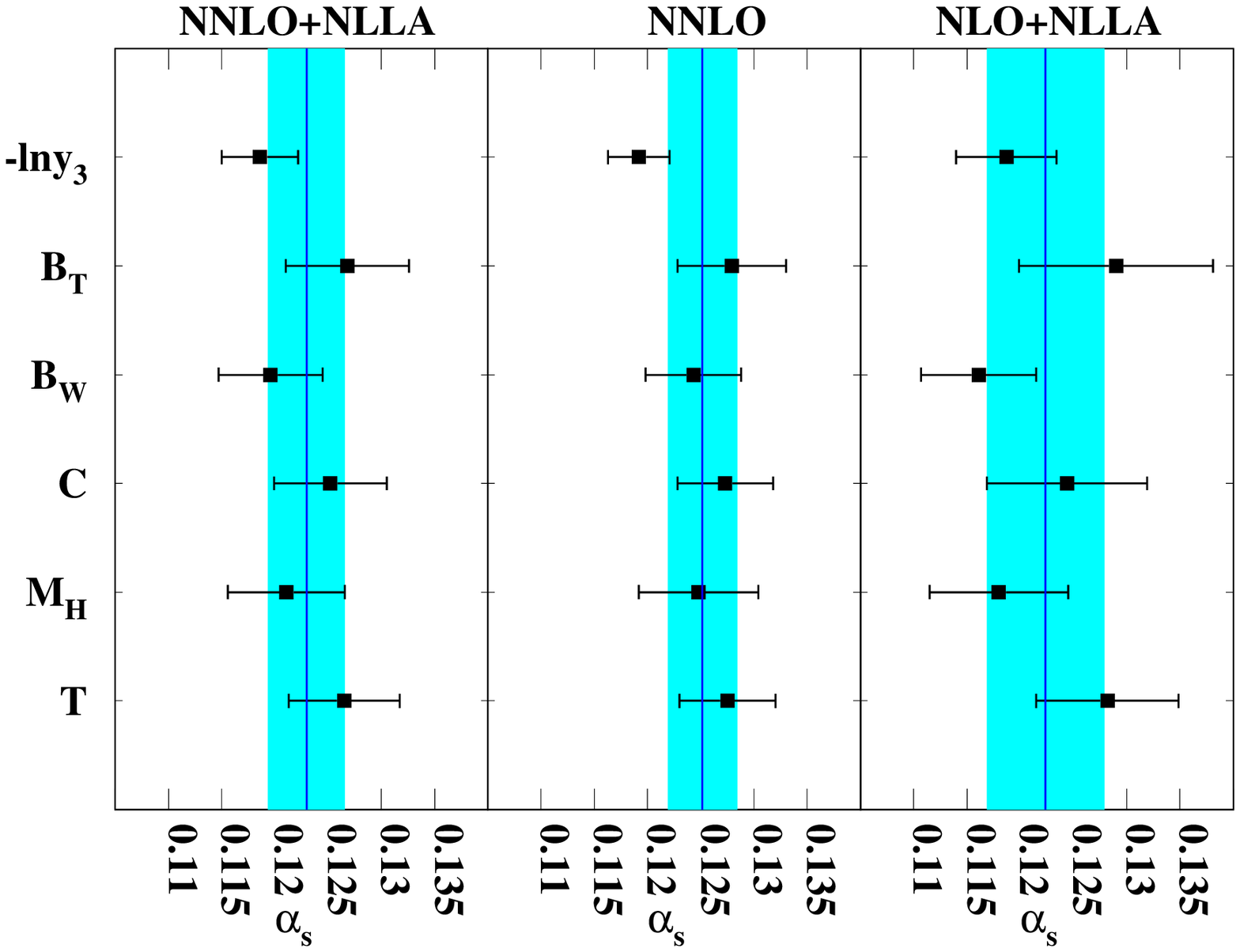}
\caption{\small The measurements of the strong coupling constant
$\alpha_s$ for the six event shapes, at $\sqrt{s}=\Mz$, when using
QCD predictions at different approximations in perturbation theory.
The shaded area corresponds to the total uncertainty, as in Fig.\
\ref{fig:comb_run}.} \protect\label{fig:scatter} }


\section{Systematic studies}
\label{sec:studies}

\subsection{$\ln R(\mu)$-matching scheme}

As described in section \ref{sec:theory}, we have computed the
two-loop terms proportional to the renormalisation scale in the
resummation and matching functions (eq. \ref{logRmumatching}) and
recomputed the theoretical error in the new matching scheme, which
we call the $\ln R(\mu)$-scheme. It is important to note that this
new matching scheme does not affect the central values of the
individual fit results, since the scale $x_\mu = x_L = 1$ is used.
Only the perturbative uncertainty will be changed because of the
different scale dependence. However, since this uncertainty enters
as a weight in the combination procedure, and different event shapes
display different scale dependence with the $\ln R(\mu)$-matching
scheme, the central values also change in the combined results. The
results for the LEP1 centre-of-mass energy are given in
Table~\ref{tab:logrmu} for all six variables, whereas the
combination of all variables and energies is listed in
Table~\ref{tab:comblogrmu}.
 The corresponding uncertainty bands are shown in Fig.~\ref{fig:bandlogrmu} for thrust and the
 three-jet transition variable. It can be seen that in this modified matching scheme the renormalisation
 scale and $x_L$ dependence are very strongly reduced, leading to a more precise
 \as{}\ determination. However, given the fact that for a consistent analysis the full NNLLA
 calculation should be matched to the NNLO prediction, we prefer to quote the values
 obtained with the standard $\ln R$-matching as our main result.

\FIGURE{
\begin{tabular}{lr}
\hspace*{-0.5cm}\includegraphics[width=7.7cm]{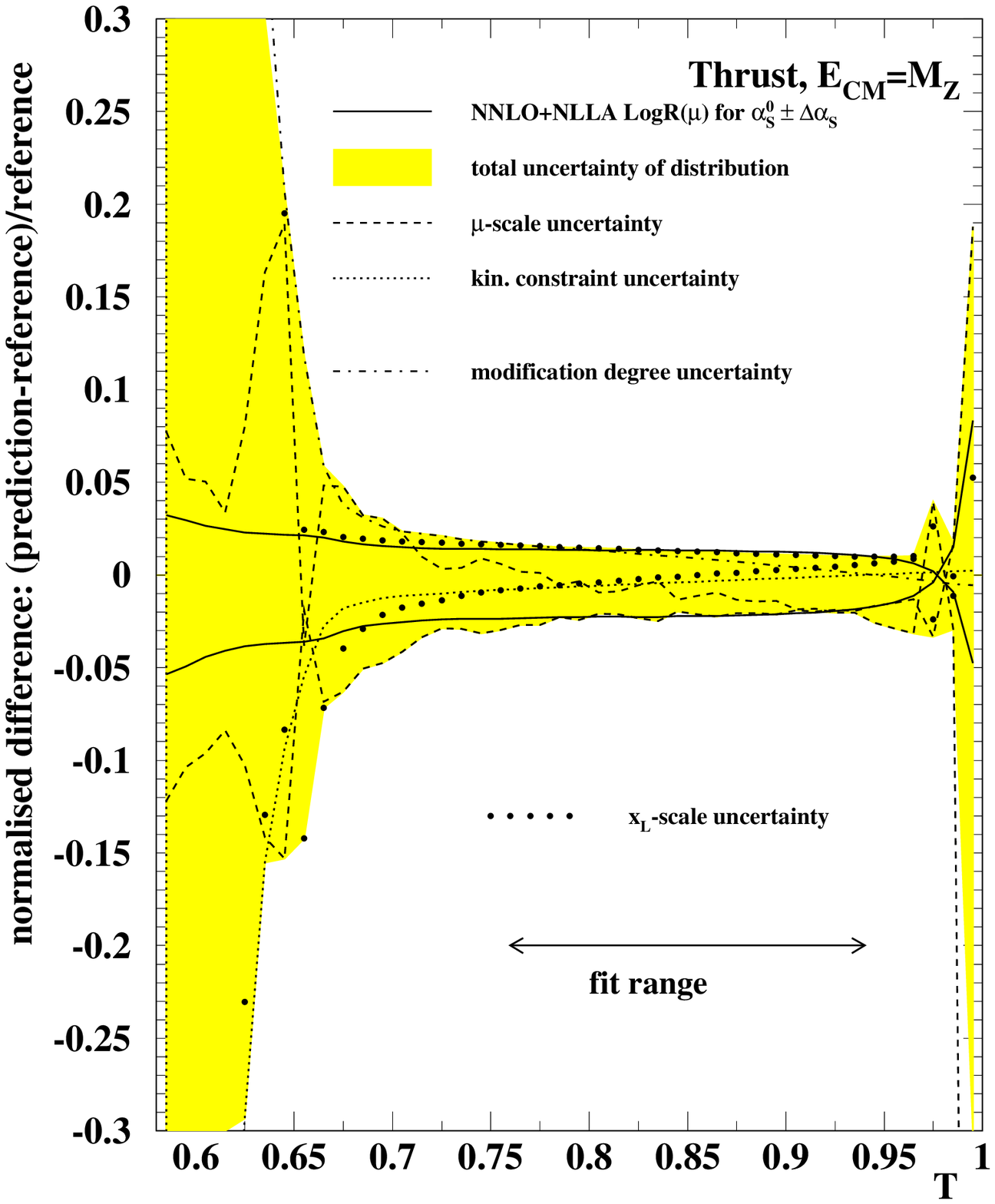} &
\hspace*{-0.5cm}\includegraphics[width=7.7cm]{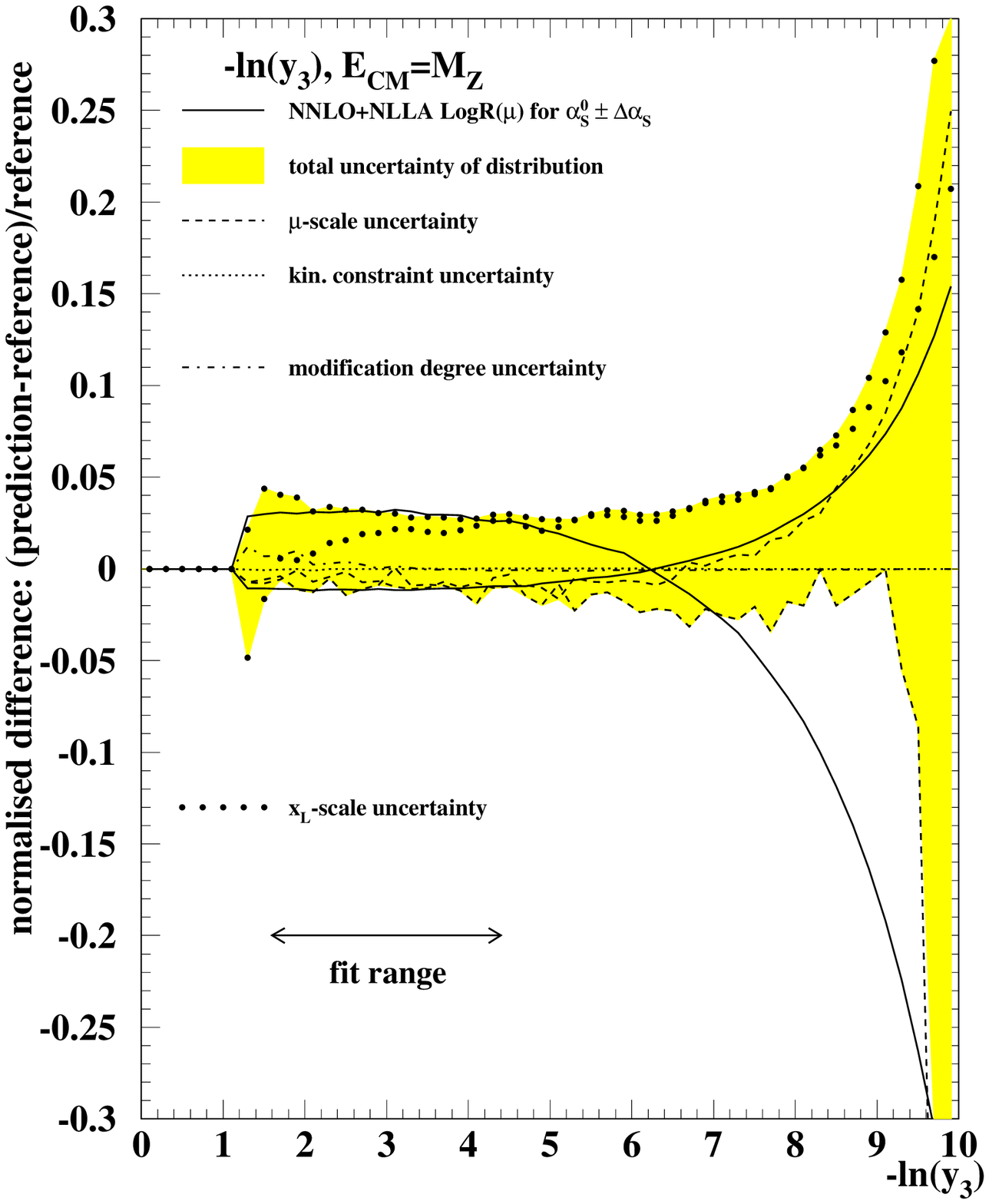}
\end{tabular}
\caption{\small  Theoretical uncertainties for the distributions of
thrust (left) and the three-jet transition variable (right) at LEP1, using the
$\ln R(\mu)$-matching scheme at NNLO+NLLA.}
\protect\label{fig:bandlogrmu}
}


\subsection{Normalisation and quark mass effects}

In our nominal analysis the theoretical prediction is normalised to
the total hadronic cross section, taking properly into account the
production of massive b-quarks. Furthermore, mass corrections are
applied for the fixed-order coefficients at leading and
next-to-leading order. In order to study the impact of different
normalisation and mass correction schemes the analysis has been
repeated with alternative options, as summarised in
Table~\ref{tab:norm} for the LEP1 data. The observed
differences when using either the massive or
massless hadronic cross section as normalisation 
are rather minor (first and second
row in Table~\ref{tab:norm}). The alternative approach to applying
the exact correction $\sigma_0/\sigma_{\mathrm{had}}$, namely
expanding this ratio and correspondingly changing the fixed order
coefficient functions $B\to {\bar B}$ and $C\to {\bar C}$, has been
adopted in previous publications. In this case the results are
lowered by about 0.5$\%$ (third row in Table~\ref{tab:norm}). For
completeness, we also give the results obtained with massless
coefficients throughout (fourth row in Table~\ref{tab:norm}). This
again lowers  the result for $\asmz$ by 0.5$\%$ at LEP1, but has
almost no effect at LEP2. The last two rows give the results obtained with 
different values for the b-quark pole mass, which we use to derive the uncertainty 
for the mass correction procedure. 


\subsection{Combination method}

Our nominal combination procedure is based on weighted averages with weights
proportional to the total significance i.e. $\propto$ $1/\sigma_{\mathrm{tot}}^2$,
where $\sigma_{\mathrm{tot}}$ is the total uncertainty for an individual measurement, thus including
the perturbative error in the weight calculation. However, it can  be expected
that the theoretical uncertainty for a given observable is highly correlated
between different energies, the main de-correlation effect being related to the different fit ranges.
In contrast, the theoretical uncertainties of different variables at the same energy are clearly
less correlated, since missing higher-order contributions are likely to be different.
Therefore it is instructive to study the stability of the combination method by using only the
largely uncorrelated uncertainty component as weight when combining the results from different
energies, while for the first-step combination of different variables at the same energy the
theoretical uncertainties are still included. As a result of such a procedure the importance
of the statistical error is significantly enhanced, leading to a reduced weight of the LEP2 data
with respect to LEP1.

In Table~\ref{tab:weights}, the newly obtained weights are compared  to the nominal weights
for the combination of measurements at different energies, and the resulting combined measurements
are given in Table~\ref{tab:combwgt}. As anticipated, now the overall combination as well as the resulting
theoretical uncertainty are dominated by the 
LEP1 data, while almost no difference is observed when combining only the LEP2 energy points.

\subsection{Hadronisation corrections from NLO+LL event generators}

In recent years substantial progress has been achieved in the development of 
modern Monte Carlo event generators targeted in particular towards the 
LHC era and often implemented in object oriented C++ frameworks. Compared to 
the legacy generators used in the LEP era, the new programs include in 
part NLO corrections matched to  parton showers 
at leading logarithmic accuracy (LL) for various processes. 
Here we use HERWIG++\,\cite{HERWIG++} version 2.3 
for our investigations, which is based on ThePEG \cite{ThePEG}, 
a general framework for implementing Monte Carlo generator classes. 
The nominal version for HERWIG++ uses a 
LO+LL configuration which features matrix element corrections for the matching 
of the hard scattering process to the parton showers. Furthermore, two schemes for the implementation 
of NLO corrections, namely the MCNLO \cite{MCNLO} and POWHEG \cite{POWHEG} schemes, are available\footnote{We use 
the notation MCNLO for the {\it method}, while MC@NLO denotes the {\it program}.}. 
The actual implementations of the general NLO to LL matching prescriptions are given in 
Ref.\ \cite{hw++MCNLO} for MCNLO and in Ref.\ \cite{hw++POWHEG} for POWHEG. Technically, a simulation 
at NLO+LL is obtained in two steps, where first the NLO partonic configurations are generated \cite{MCPWNLO} and second 
these events are passed to HERWIG++ to simulate the parton shower, 
hadronisation and resonance decays.  

It should be noted that the nominal  LO+LL version with matrix element corrections of HERWIG++ has 
been extensively tuned to a variety of experimental data, including 
jet rates, event shapes and particle multiplicities in \epm\ annihilations from LEP and heavy flavour data from the B-factories \cite{HW++tuning}, in order 
to obtain the best possible set of parameters. However, according to \cite{HW++tuning}, the quality of this fit, with an overall $\chi^2$ 
per degree of freedom between 5 and 6, is limited, and this is, to some extent, 
related to a slightly overestimated amount of gluon radiation in the parton shower. 
It can not be expected that this tuning is optimal for the NLO+LL versions of HERWIG++. A re-tuning 
of the parameters using MCNLO and POWHEG is beyond the scope of this paper, but we used 
parameters suggested in Ref.\ \cite{MCNLOtuning} for MCNLO and checked a few of the main parameters for 
POWHEG using the ALEPH event shapes\,\cite{ALEPH-qcdpaper}. To this end four parameters 
were individually varied in a simplified grid-search procedure in which for each configuration the $\chi^2$ with respect to the full range of 
the six event-shape distributions from ALEPH at LEP1 was recorded. This resulted in an improved 
description of the distributions studied here, presumably at the expense of other distributions 
included in the more general tuning of HERWIG++.
 
The following parameters were determined according to this procedure for POWHEG:
\textbf{AlphaMZ} = 0.134, \textbf{cutoffKinScale} = 2.7, \textbf{PSplitLight} = 1.1 and \textbf{PwtDIquark} = 0.6.
The meaning of these parameters can be inferred from \cite{HERWIG++}. In addition, the partonic 
configuration was generated with a value of $\Lambda =$ 170 MeV to set the scale for the hardest gluon emission
according to the evolution in the $\overline{\mathrm{MS}}$-scheme, yielding $\asmz$=0.11.  

We  compare in a first step the prediction for the 
event shape distributions of HERWIG++ to both the high precision data at LEP1 
from ALEPH and the predictions from the legacy generators PYTHIA, HERWIG and ARIADNE. We recall that the latter have 
all been tuned to the same global QCD observables measured by ALEPH \cite{aleph_mega} at LEP1, which included event-shape variables 
similar to the ones analysed here.  
In Fig.\ \ref{fig:hw++tuning} the generator 
predictions for thrust, $-\ln(y_3)$ and the total and wide jet broadenings are compared to the ALEPH data. 

\FIGURE{
\begin{tabular}{lr}
\hspace*{-0.4cm}\includegraphics[width=7.0cm]{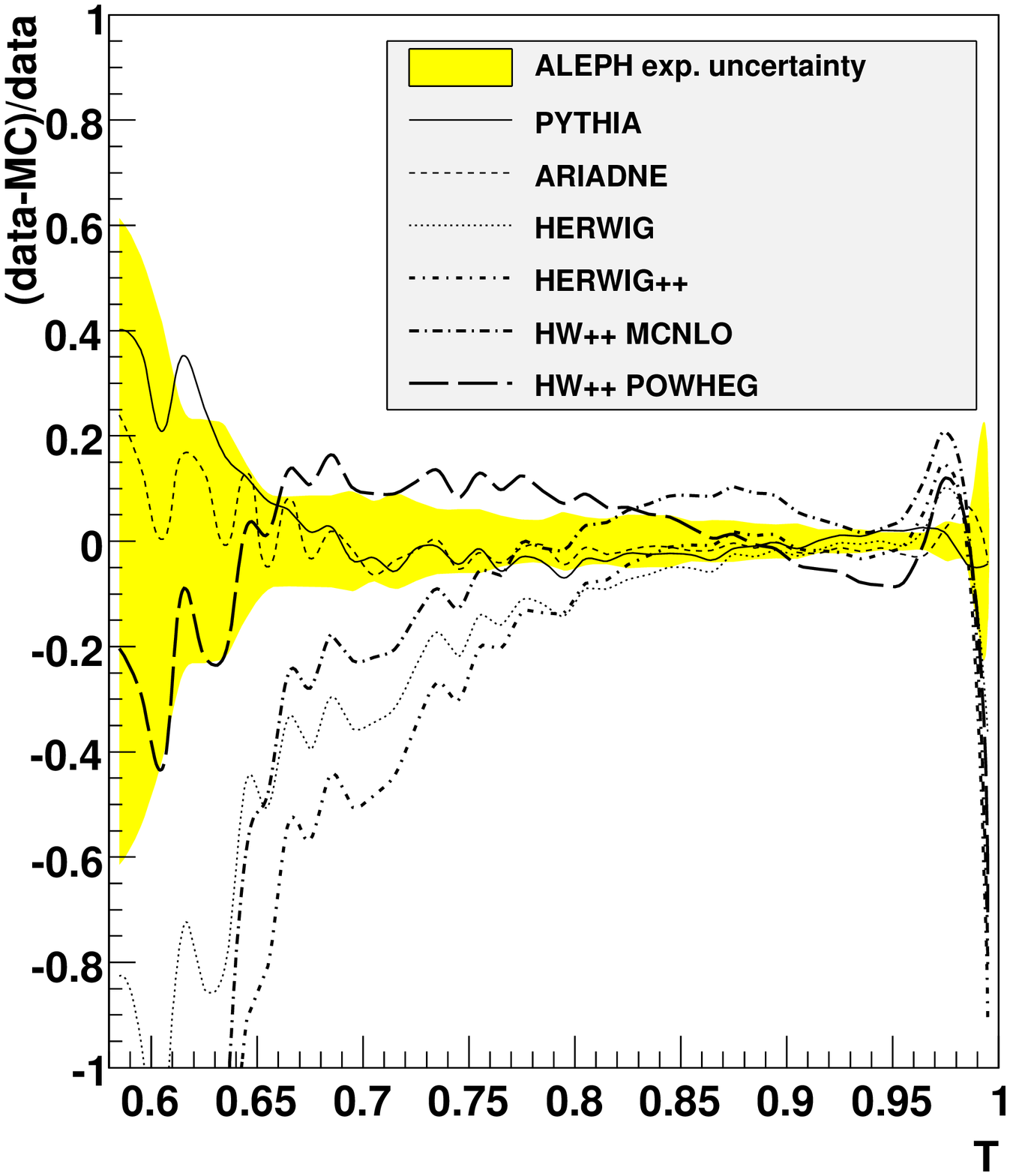} &
\hspace*{-0.5cm}\includegraphics[width=7.0cm]{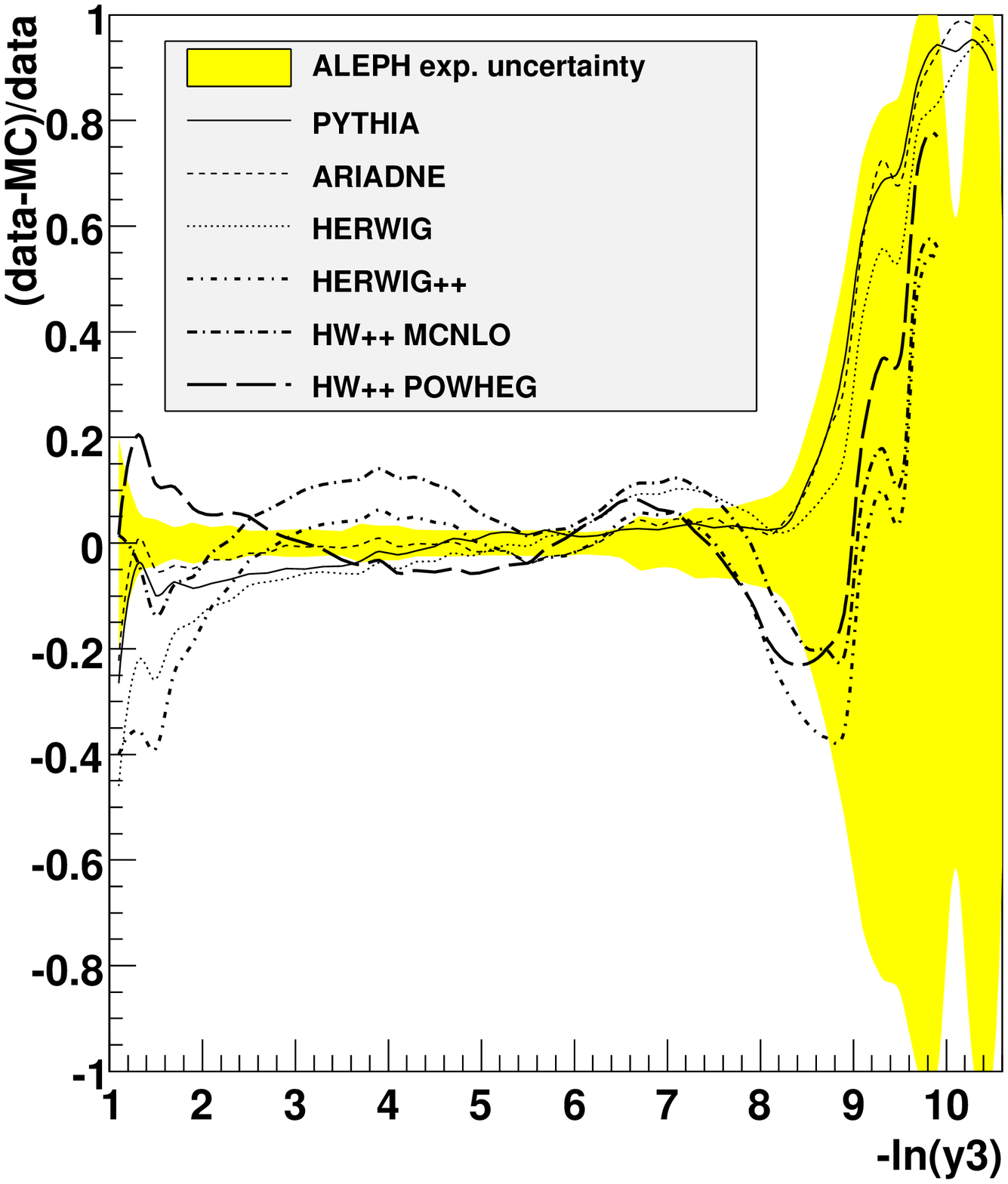} \\
\hspace*{-0.4cm}\includegraphics[width=7.0cm]{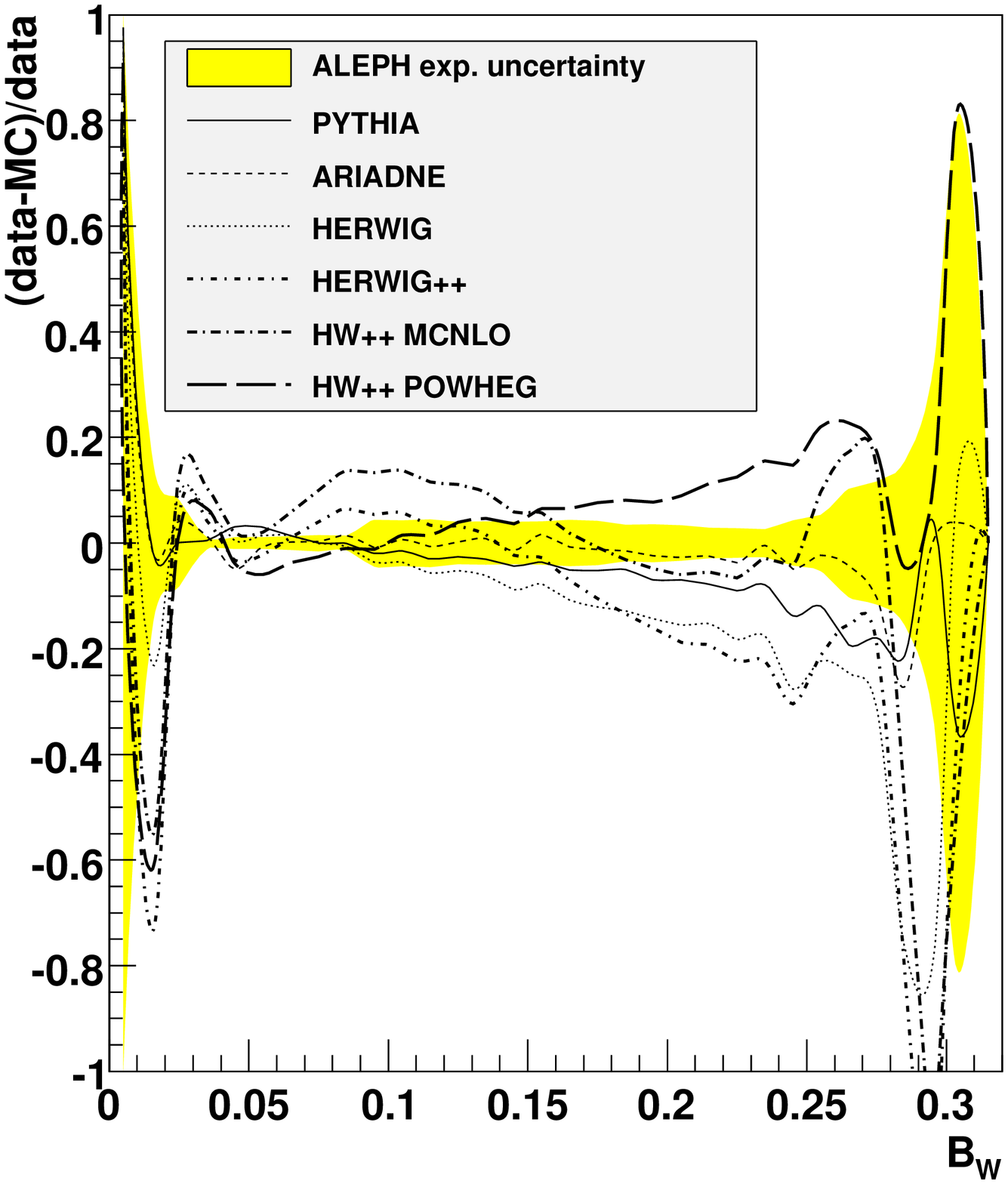} &
\hspace*{-0.5cm}\includegraphics[width=7.0cm]{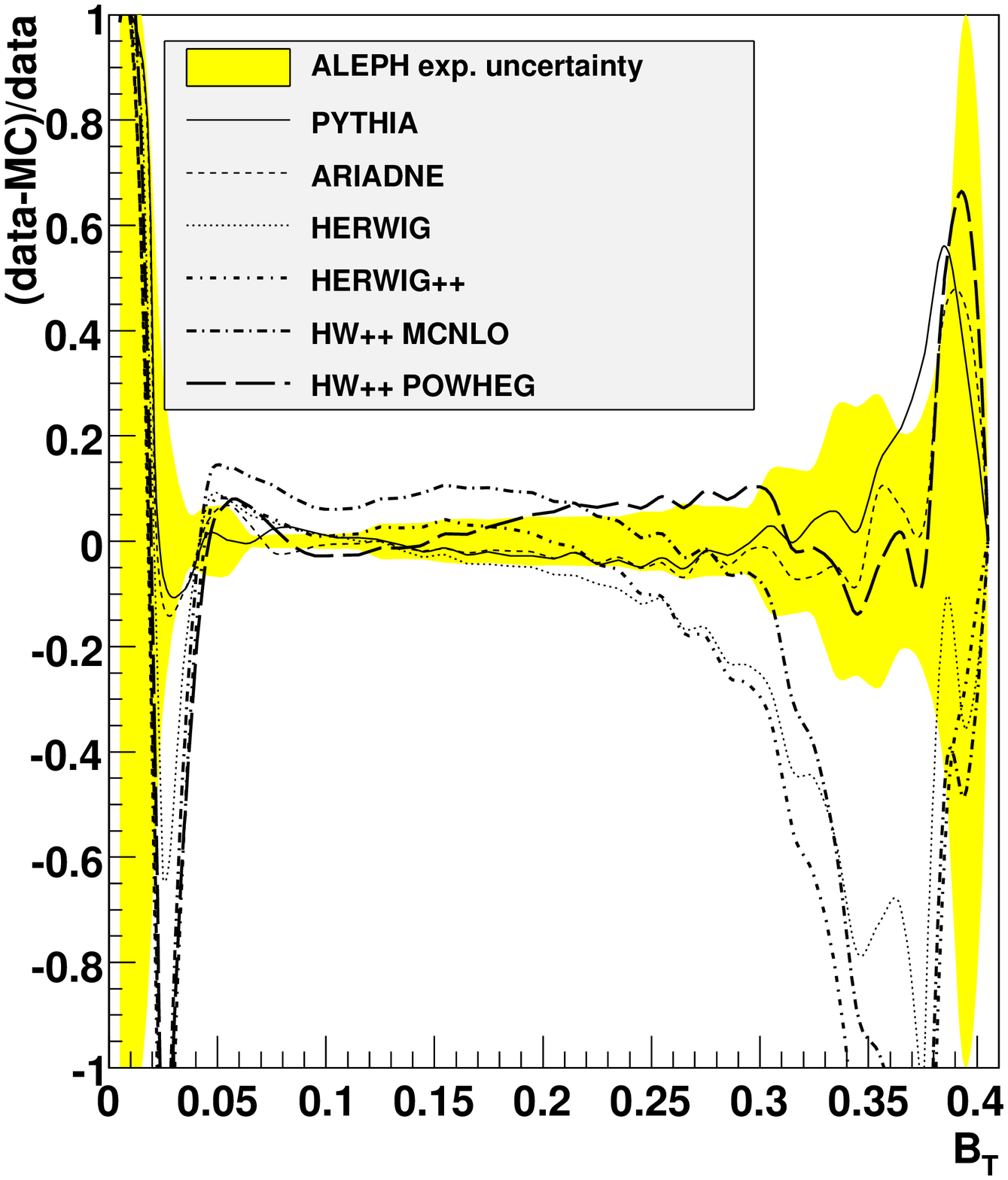}
\end{tabular}
\caption{Residuals of hadron level Monte Carlo predictions with respect 
to the ALEPH data. The shaded area indicates the experimental uncertainty.}
\protect\label{fig:hw++tuning}}
In general it appears that the shape of HERWIG++ is similar to both HERWIG++ with MCNLO and HERWIG, but all differ in 
normalisation. A better description is obtained using HERWIG++ with POWHEG. PYTHIA and ARIADNE yield by far the best description.  
 To quantify the performance of the generators,  in Table \ref{tab:hw++tune} we have compiled  
the $\chi^2$ of their predictions with respect to the event shapes studied  at LEP1, including the 
experimental systematic uncertainty. A complete re-tuning of the HERWIG++ parameters to the same data used to 
tune PYTHIA, HERWIG and ARIADNE would likely improve their performance.      

To investigate the origin of the observed differences between the 
generators, it is instructive to consider the parton-level predictions and 
the hadronisation corrections separately. 
The parton level predictions from the generators are calculated with final state partons at the end 
of the parton shower. These are compared to the complete NNLO+NLLA calculation in Fig.~\ref{fig:hw++nnlonlla}. 
For thrust and in particular the total jet broadening a reasonable agreement between NNLO+NLLA and 
HERWIG++ with POWHEG, as well as a fair agreement with PYTHIA and 
ARIADNE is observed, while other HERWIG variants show a clear deviation. 
For $-\ln(y_3)$ and the wide jet broadening all legacy generators 
provide a satisfactory description and HERWIG++ based predictions exhibit some systematic differences in shape.
It is worth noting that for thrust and the total jet broadening, the PYTHIA prediction overestimates the 
NNLO+NLLA  calculation by about $10\%$, with the shape in reasonable agreement, whereas a better
agreement is seen for the wide jet broadening and $-\ln(y_3)$.

\FIGURE{
\begin{tabular}{lr}
\hspace*{-0.4cm}\includegraphics[width=7.0cm]{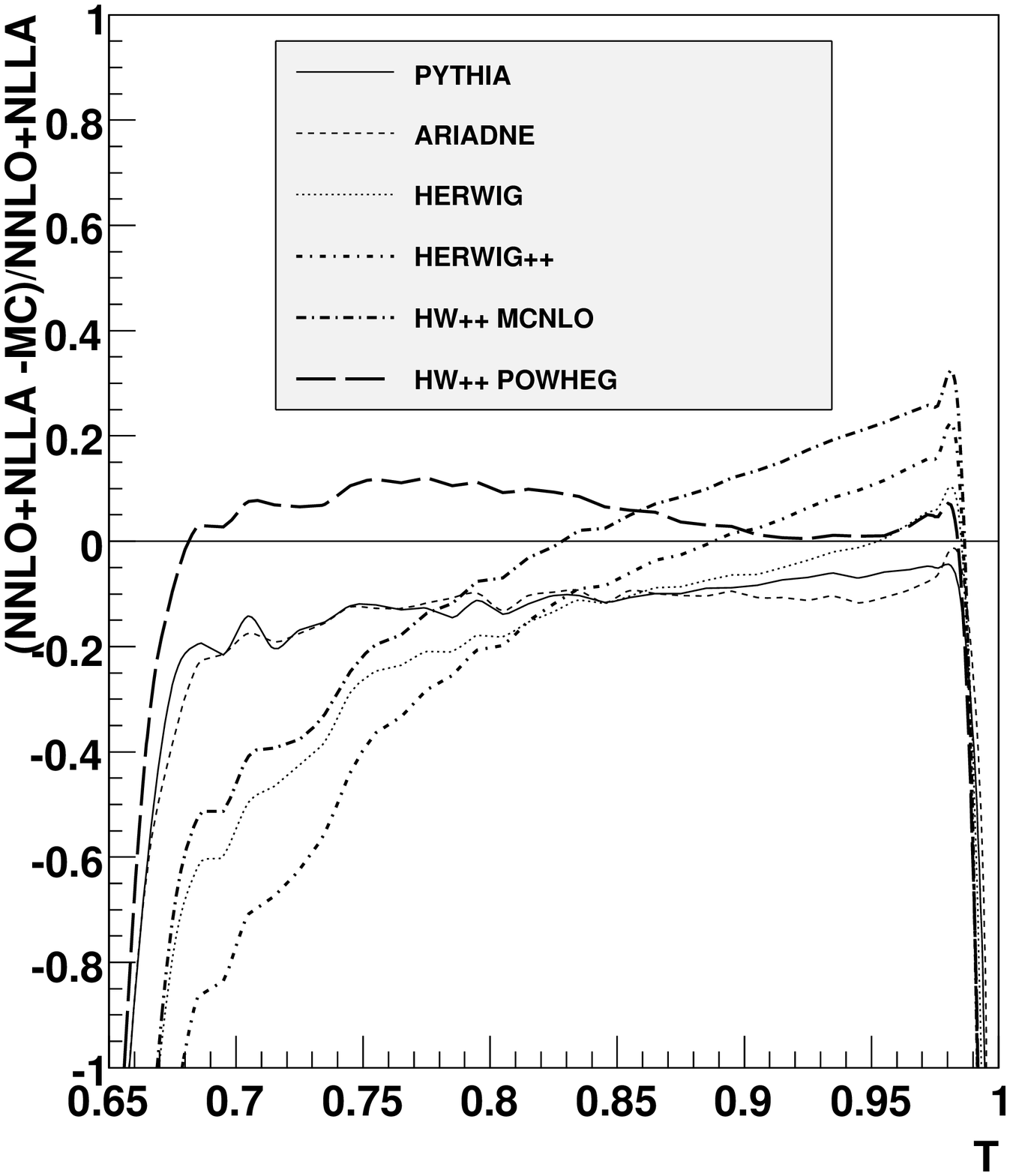} &
\hspace*{-0.5cm}\includegraphics[width=7.0cm]{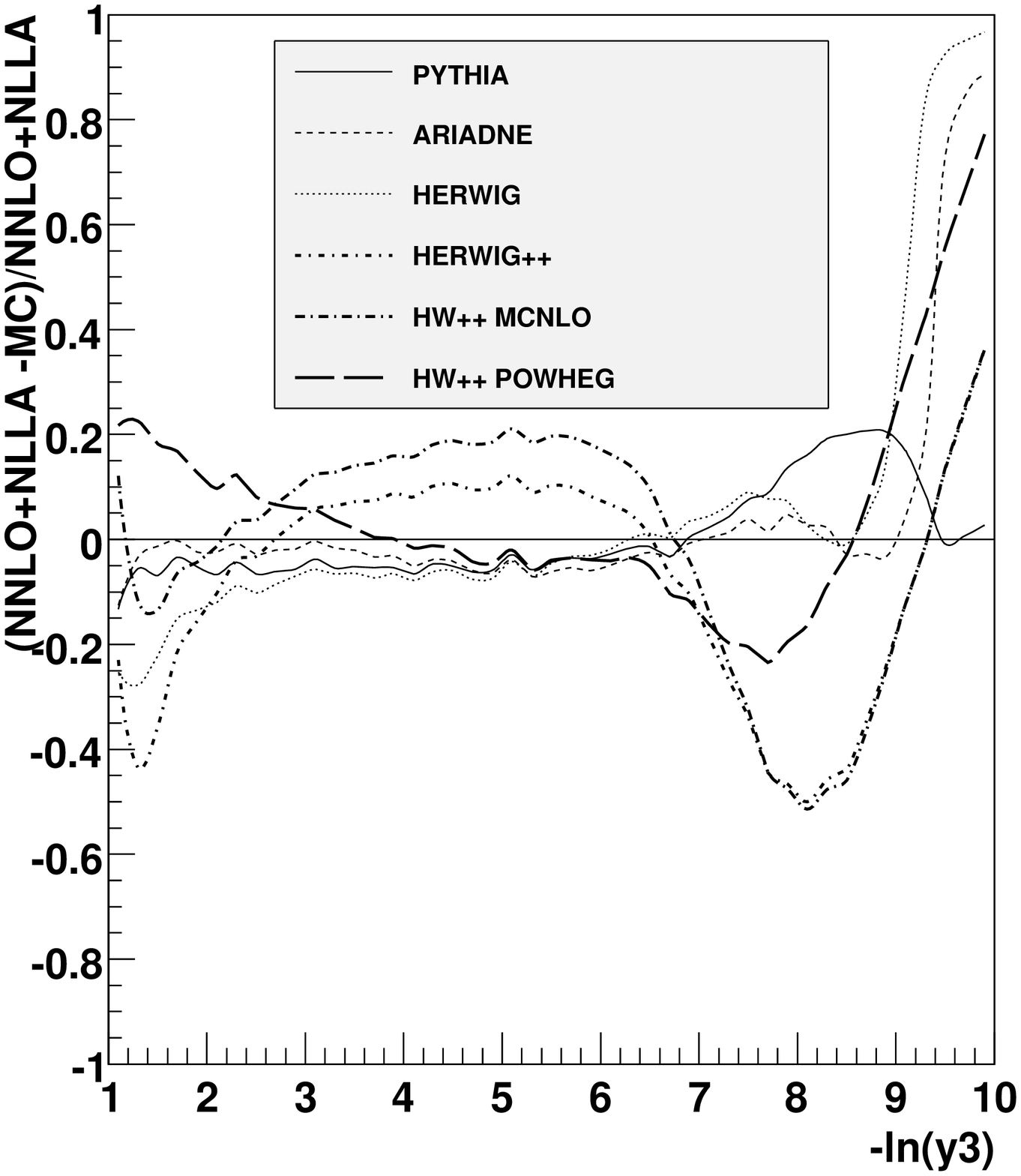}\\ 
\hspace*{-0.4cm}\includegraphics[width=7.0cm]{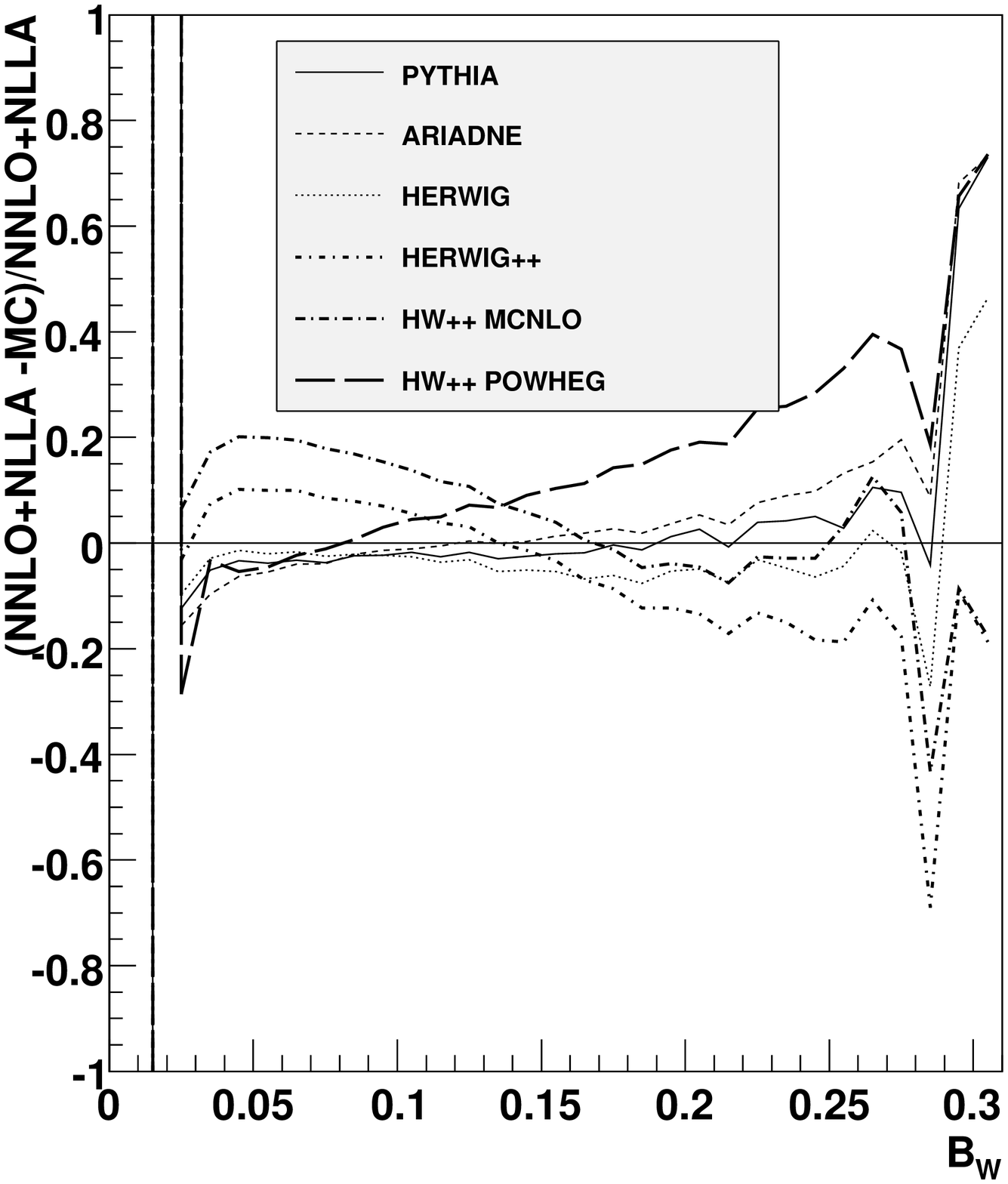} &
\hspace*{-0.5cm}\includegraphics[width=7.0cm]{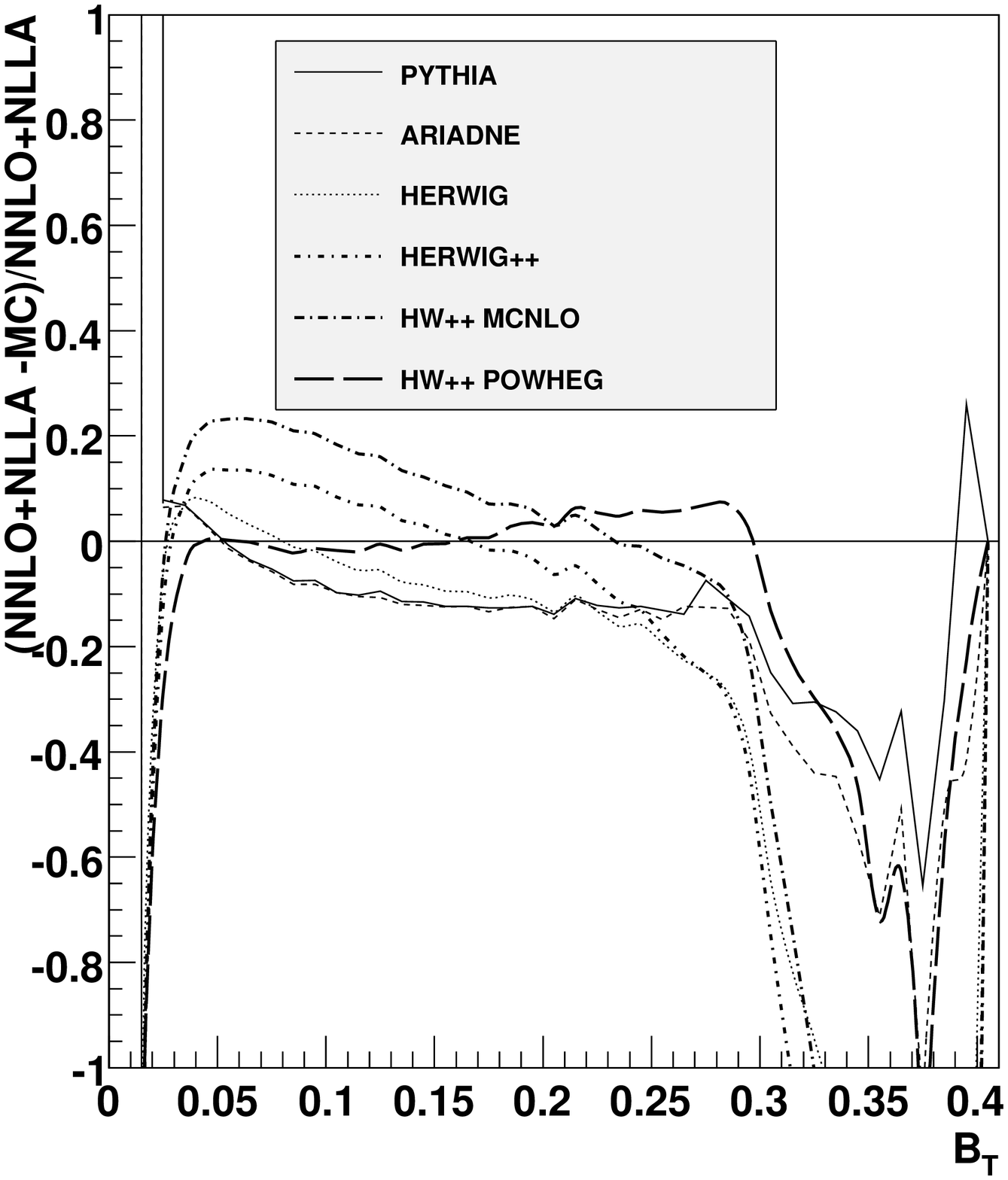}
\end{tabular}
\caption{Normalised residuals with respect to the NNLO+NLLA calculation of parton level predictions obtained from different Monte Carlo generators.}
\protect\label{fig:hw++nnlonlla}}

The hadronisation corrections to be used in the fits to the data are shown in Fig.\ \ref{fig:hw++hadcor}. 
HERWIG++ with POWHEG yields a similar shape as the legacy programs, but differs in the normalisation.
The other HERWIG++ predictions differ most notably in shape from the former. 
In Table \ref{tab:hw++fits} the fit results obtained with all generators for hadronisation corrections are given. 
In most cases, the fits based on HERWIG++ and HERWIG++ with MCNLO are significantly worse than for the other generators, 
but for individual variables like the wide jet broadening an opposite behaviour is observed. The fit quality 
of HERWIG++ with POWHEG is similar to the outcome of the legacy generators. 
Given the similar shape but different 
normalisation of HERWIG++ with POWHEG, the resulting values of \as\, are significantly lower, overall by 3\,$\%$.     
\FIGURE{
\begin{tabular}{lr}
\hspace*{-0.4cm}\includegraphics[width=7.0cm]{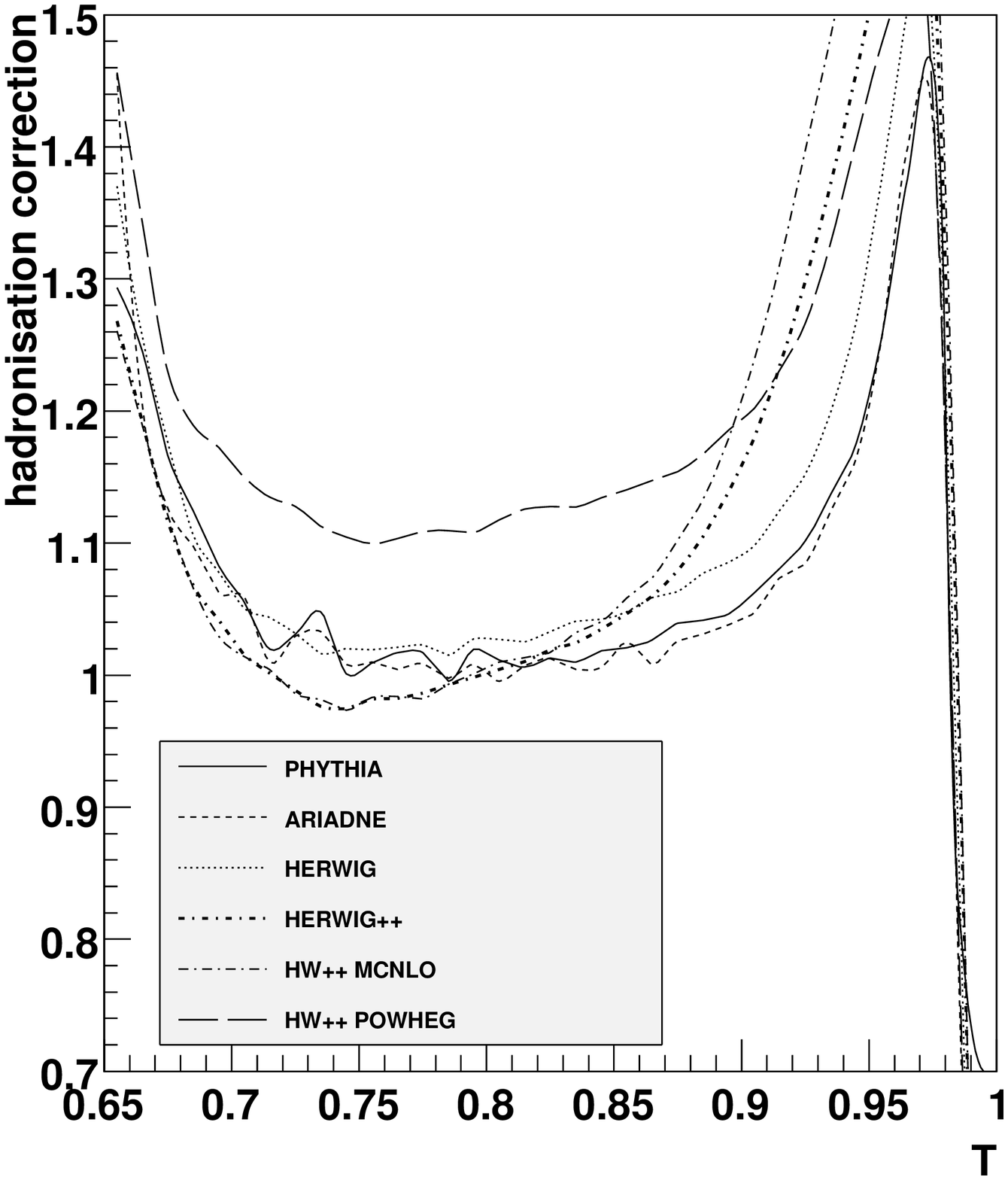} &
\hspace*{-0.5cm}\includegraphics[width=7.0cm]{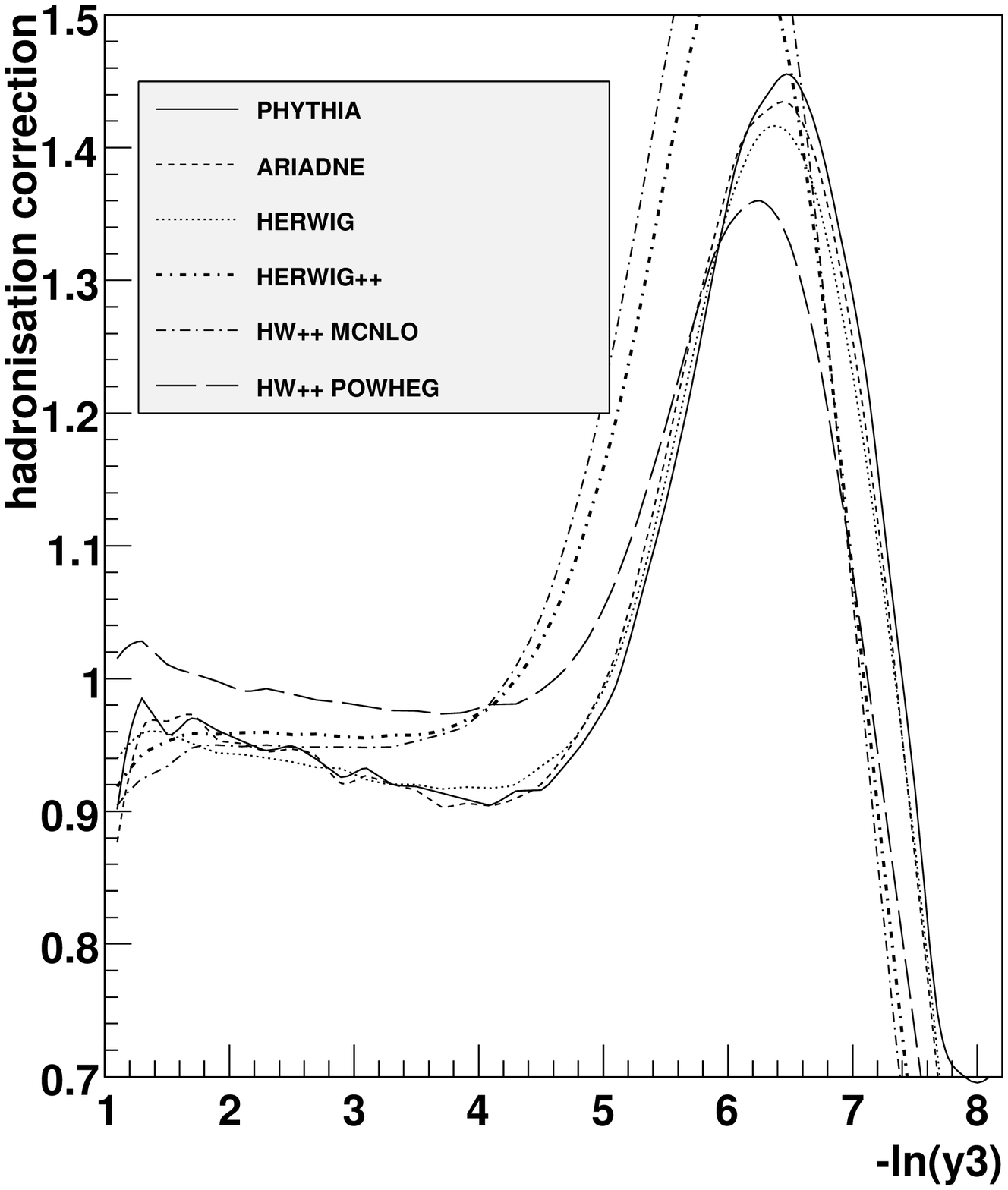} \\
\hspace*{-0.4cm}\includegraphics[width=7.0cm]{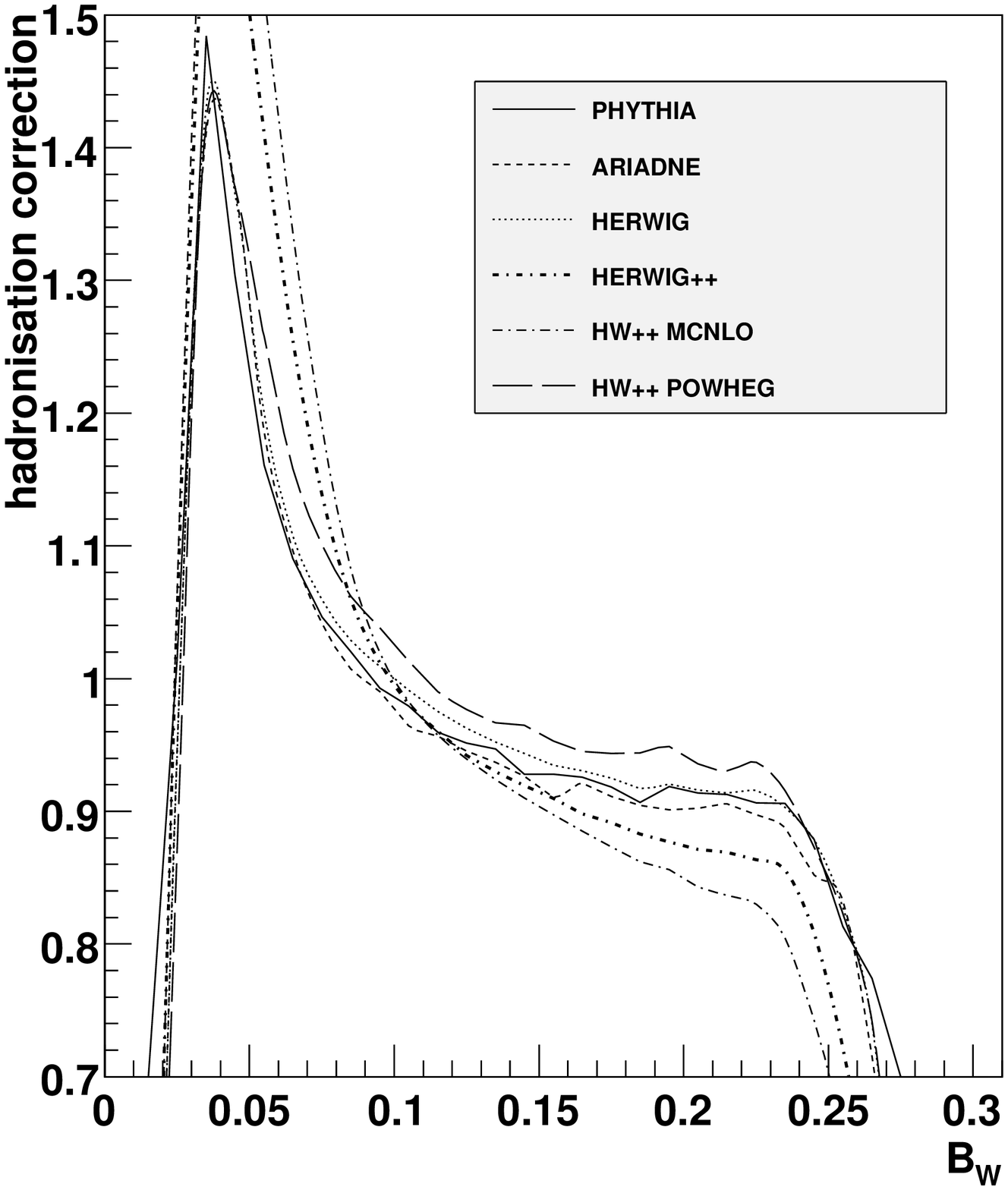} &
\hspace*{-0.5cm}\includegraphics[width=7.0cm]{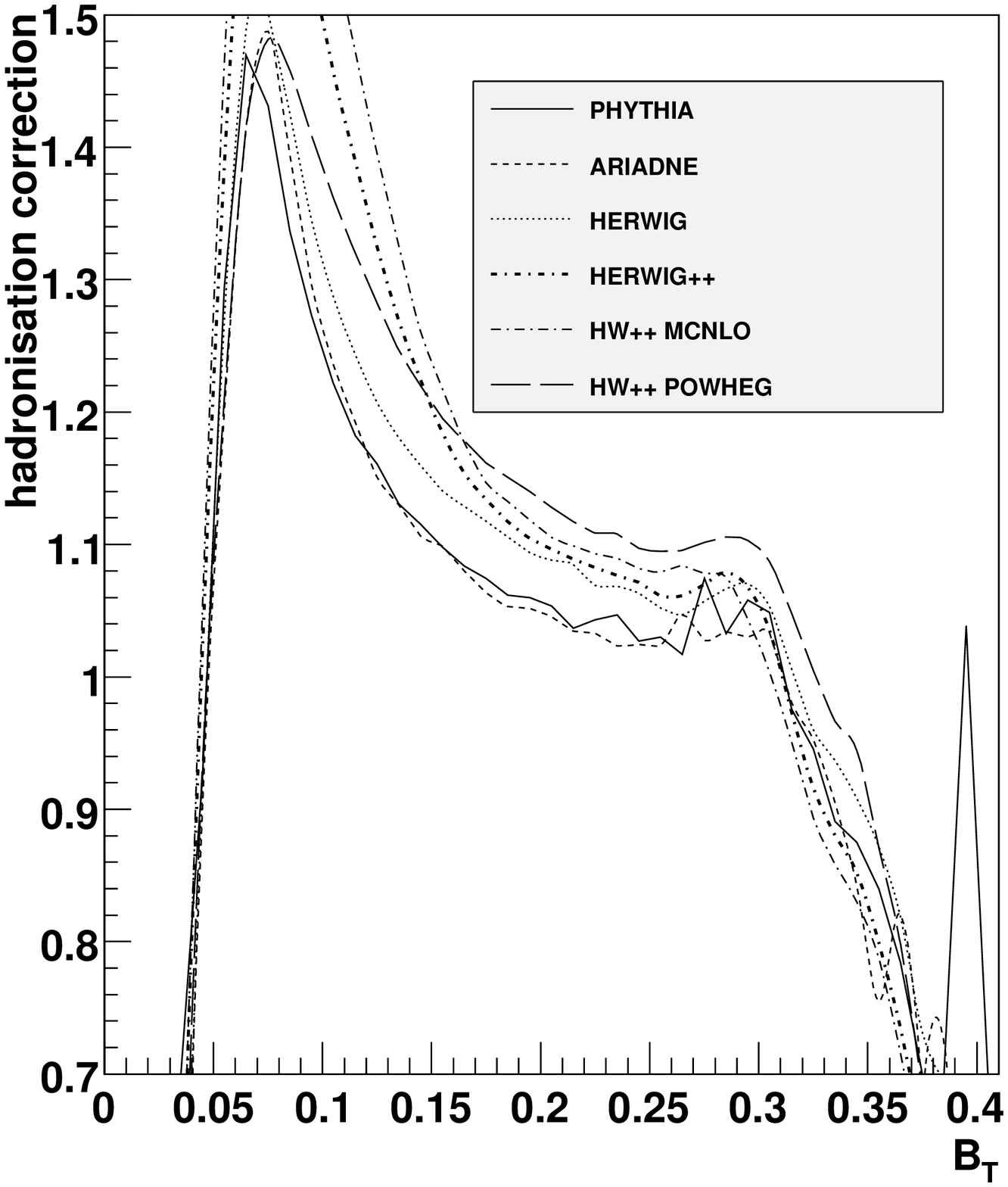}
\end{tabular}
\caption{Hadronisation corrections (ratio of hadron to parton level predictions)  obtained from different Monte Carlo generators.}
\protect\label{fig:hw++hadcor}}


\section{Discussion and Conclusions}
\label{sec:discussion}

We have performed a determination of the strong coupling constant
\as{}\ from event-shape data measured by the ALEPH
collaboration~\cite{ALEPH-qcdpaper}, based on the perturbative QCD
results at  next-to-next-to-leading order (NNLO) matched to
resummation in the next-to-leading-logarithmic approximation
(NLLA)~\cite{Gehrmann:2008kh}.

Comparing our results to both the fit using purely fixed-order
NNLO predictions~\cite{Dissertori:2007xa}
and  the fits based on earlier NLLA+NLO calculations~\cite{ALEPH-qcdpaper},
we make the following observations:
\begin{itemize}
\item The central value obtained by combining the results for six event-shape variables
and the LEP1 and LEP2 centre-of-mass energies,
\begin{center}
    $\asmz = 0.1224
    \;\pm\; 0.0009\,\mathrm{(stat)}
    \;\pm\; 0.0009\,\mathrm{(exp)}
    \;\pm\; 0.0012\,\mathrm{(had)}
    \;\pm\; 0.0035\,\mathrm{(theo)}$,
\end{center}
is slightly lower than the central value of 0.1228 obtained from fixed-order NNLO only,
and slightly larger than the NLO+NLLA results. We note that in this analysis
an improved normalisation to the total hadronic cross section has been used, which leads to minor deviations to previously reported results.

The fact that the central value is almost identical to
the purely fixed-order NNLO result could be anticipated from the
findings in Ref.~\cite{Gehrmann:2008kh}. There it is shown that
in the three-jet region, which provides the bulk of the fit range,
the matched NLLA+NNLO prediction is very close to the fixed-order NNLO calculation.

\item The dominant theoretical uncertainty on  \asmz, as estimated from
scale variations, is reduced by 20\% compared to NLO+NLLA.
However, compared to the fit based on purely fixed-order
NNLO predictions, the perturbative uncertainty is {\it increased} in the NNLO+NLLA fit.
The reason is that in the two-jet region the NLLA+NLO and NLLA+NNLO
predictions agree by construction, because the matching suppresses any
fixed order terms. Therefore, the renormalisation scale uncertainty
is dominated by the next-to-leading-logarithmic
approximation  in this region, which results in a larger overall
scale uncertainty in the \as{}\ fit.

\item As already observed for the fixed-order NNLO results,
the scatter among the values of \asmz\ extracted from the six
different event-shape variables is smaller than in the NLO+NLLA
case.

\item The matching of NLLA+NNLO introduces a mismatch in the cancellation
of renormalisation scale logarithms, since the NNLO expansion fully
compensates the renormalisation scale dependence up to two loops,
while NLLA only compensates it up to one loop. In order to assess
the impact of this mismatch, we introduced the $\ln R(\mu)$ matching
scheme, which retains the two-loop renormalisation terms in the
resummed expressions and the matching coefficients. In this scheme,
a substantial reduction of the perturbative uncertainty from
$\pm0.0035$ (obtained in the default $\ln R$-scheme) to $\pm 0.0022$
is observed, which might indicate the size of the ultimately
reachable precision for a complete NNLO+NNLLA calculation including
the currently unknown resummed function $g_3$ for all shape
variables. 
Although both schemes are in principle on the same
theoretical footing, it is the more conservative error
estimate obtained in the $\ln R$-scheme which should be taken as the
nominal value, since it measures the potential
impact of the yet uncalculated finite NNLLA-terms.

\item Bottom quark mass effects, which are numerically significant mainly at the LEP1 energy, 
were included through to NLO. Compared to a purely massless evaluation of the distributions, the inclusion of these mass effects enhances $\asmz$ by 0.8\%. Compared to the previously used expansion of the mass corrections, an enhancement of 0.4\% is observed.

\item The averaging of $\asmz$ values obtained at the various LEP1 and LEP2 energies weights the different measurements by their total uncertainties. Excluding the error on the perturbative prediction from this weighting enhances the importance of the very precise LEP1 data over the LEP2 data, and yields an $\asmz$ value which is lower than our default result by only 0.2\%, thereby demonstrating the very good consistency of the LEP1 and LEP2 results.

\item We have investigated hadronisation corrections obtained from NLO+LL parton shower simulation using HERWIG++ 
with two different schemes. Results for \as\, based on corrections from HERWIG++ with POWHEG are slightly 
lower than with nominal corrections from PYTHIA. Comparing hadron level predictions with data reveals that 
HERWIG++ with POWHEG yields an improved prediction over HERWIG, HERWIG++ and HERWIG++ with MCNLO, but does not 
reach the same level of agreement as PYTHIA and ARIADNE. 
Further, we observe a certain discrepancy between 
MCNLO and POWHEG, which might indicate unresolved tuning issues. 
Therefore, while the first studies with 
HERWIG++ look rather promising, we retain for the time being PYTHIA as generator for our nominal result.      

\item From the study of hadronisation corrections we also make the following important observation. It appears that there are two
        ``classes'' of variables.  The first class contains thrust, C-parameter and total jet broadening, whereas the second
           class consists of the heavy jet mass, wide jet broadening and the two-to-three-jet transition parameter  $-\ln y_3$.
           For the first class, using the standard hadronisation corrections from PYTHIA, we obtain $\asmz$ values around
           $0.125 - 0.127$, some $5\%$ higher than those found from the second class of variables. In a study of 
            higher moments of event shapes \cite{Ridder:2009dp}, 
            indications were found that variables from the first class still exhibit sizable
             missing higher order corrections, whereas the second class of observables have a better perturbative stability.
              In this paper, from Fig.\ \ref{fig:hw++nnlonlla}, we observe that this first class
              of variables gives a parton level prediction with PYTHIA, which is about $10\%$ higher than the NNLO+NLLA
              prediction. The PYTHIA curve is obtained with tuned parameters, where the tuning to data had been performed
             at the hadron level. Indeed, this tuning results in a rather large effective coupling in the parton shower, which might
              partly explain the larger parton level prediction of PYTHIA. However, since the tuning has been performed
             at hadron level, this implies that the hadronisation corrections come out to be smaller than what would have
              been found by tuning a hypothetical Monte Carlo prediction with a parton level corresponding to the 
             NNLO+NLLA prediction. Thus, in the end, the PYTHIA hadronisation corrections, applied in the
             $\as{}$ fit, might be too small, resulting in a larger $\asmz$ value. Such problems do not appear to exist for the
             second class of variables.
             
             In summary, there are indications that the first class of event shapes still suffers from significant missing higher order
            contributions, even beyond NNLO+NLLA. This might also have led to a tuning of parton shower models which
               underestimates the hadronisation corrections for these variables, and consequently results in somewhat larger
              values of the fitted strong coupling. Since up to now the hadronisation uncertainties have been estimated
                from the differences of parton shower based models, tuned to the data, it is likely that for these event shapes
               the uncertainties were underestimated and not able to account for a possible systematic shift.

\end{itemize}
In future work it would be interesting to investigate the
effect of NNLLA resummation terms for all six event shapes,
of electroweak corrections, of quark mass effects beyond NLO
and of non-perturbative power-law corrections as well as further studies 
with HERWIG++, in particular using the newly developed improved algorithm for 
merging matrix elements with angular-ordered parton showers~\cite{HW++CKKW}.


\section*{Acknowledgements}
We wish to thank the authors of HERWIG++ for fruitful discussions on hadronisation corrections 
and O.~Latunde-Dada for valuable instructions on running the MCPWNLO interface.
GH and HS would like to thank the Institute for Theoretical Physics, 
University of Z\"urich,
for hospitality while part of this work was carried out.
EWNG gratefully acknowledges the support of the
Wolfson Foundation and the Royal Society.
This research was supported in part by the Swiss National Science Foundation
(SNF) under contracts  PP0022-118864 and 200020-117602,
 by the UK Science and Technology Facilities Council,
 by the European Commission's Marie-Curie Research Training Network under contract
MRTN-CT-2006-035505 ``Tools and Precision Calculations for Physics Discoveries
at Colliders'' and by the German Helmholtz Alliance ``Physics at the Terascale''.


\bibliographystyle{JHEP}


\renewcommand{\arraystretch}{1.1}
\renewcommand{\textfraction}{0.0001}
\TABLE{
\caption[Individual results]{\small Results for
$\alpha_s(Q)$ as obtained from NNLO+NLLA fits to distributions of
event-shape variables at $Q = \sqrt{s} =$ 91.2, 133, 161 and 172 GeV.\label{tab:indiv1}}
\begin{tabular}{|l|c|c|c|c|c|c|}\hline
\multicolumn{7}{|c|}{ $Q$ = 91.2 GeV} \\ \hline
variable        & $T$   & $-\ln y_3$ & $M_H$ & $C$ & $B_W$ & $B_T$  \\ \hline
$\alpha_s$      & 0.1265 & 0.1186 & 0.1211 & 0.1252 & 0.1196 & 0.1268 \\
stat. error     & 0.0002 & 0.0002 & 0.0003 & 0.0002 & 0.0002 & 0.0002 \\
exp. error      & 0.0008 & 0.0011 & 0.0010 & 0.0007 & 0.0007 & 0.0007 \\
pert. error     & 0.0048 & 0.0029 & 0.0033 & 0.0050 & 0.0046 & 0.0053 \\
hadr. error     & 0.0019 & 0.0017 & 0.0042 & 0.0016 & 0.0017 & 0.0022 \\
total error     & 0.0052 & 0.0036 & 0.0055 & 0.0053 & 0.0049 & 0.0058 \\ \hline
fit range       & 0.75-0.91 & 1.6-4.0 & 0.10-0.22 & 0.36-0.74 & 0.09-0.19 & 0.16-0.30 \\ \hline
\multicolumn{7}{|c|}{$Q$ = 133 GeV} \\ \hline
variable        &  $T$   & $-\ln y_3$ & $M_H$ & $C$ & $B_W$ & $B_T$  \\ \hline
$\alpha_s$      & 0.1193 & 0.1199 & 0.1149 & 0.1174 & 0.1158 & 0.1205 \\
stat. error     & 0.0047 & 0.0057 & 0.0053 & 0.0037 & 0.0032 & 0.0031 \\
exp. error      & 0.0006 & 0.0004 & 0.0012 & 0.0008 & 0.0008 & 0.0010 \\
pert. error     & 0.0039 & 0.0024 & 0.0027 & 0.0040 & 0.0038 & 0.0044 \\
hadr. error     & 0.0015 & 0.0010 & 0.0027 & 0.0012 & 0.0010 & 0.0013 \\
total error     & 0.0063 & 0.0063 & 0.0067 & 0.0057 & 0.0051 & 0.0056 \\
fit range       & 0.75-0.94 & 1.6-4.4 & 0.08-0.25 & 0.30-0.75 & 0.08-0.25 & 0.13-0.35 \\ \hline
\multicolumn{7}{|c|}{$Q$ = 161 GeV} \\ \hline
variable        &  $T$   & $-\ln y_3$ & $M_H$ & $C$ & $B_W$ & $B_T$  \\ \hline
$\alpha_s$      & 0.1172 & 0.1183 & 0.1225 & 0.1190 & 0.1186 & 0.1238 \\
stat. error     & 0.0080 & 0.0082 & 0.0072 & 0.0066 & 0.0047 & 0.0052 \\
exp. error      & 0.0006 & 0.0004 & 0.0012 & 0.0008 & 0.0008 & 0.0010 \\
pert. error     & 0.0036 & 0.0022 & 0.0025 & 0.0037 & 0.0035 & 0.0040 \\
hadr. error     & 0.0014 & 0.0007 & 0.0022 & 0.0011 & 0.0008 & 0.0010 \\
total error     & 0.0088 & 0.0085 & 0.0079 & 0.0076 & 0.0060 & 0.0067 \\\hline
fit range       & 0.75-0.94 & 1.6-4.4 & 0.08-0.25 & 0.30-0.75 & 0.08-0.25 & 0.13-0.35 \\ \hline
\multicolumn{7}{|c|}{ $Q$ = 172 GeV} \\ \hline
variable        & $T$   & $-\ln y_3$ & $M_H$ & $C$ & $B_W$ & $B_T$  \\ \hline
$\alpha_s$      & 0.1120 & 0.1095 & 0.1079 & 0.1093 & 0.1036 & 0.1108 \\
stat. error     & 0.0077 & 0.0098 & 0.0085 & 0.0063 & 0.0063 & 0.0069 \\
exp. error      & 0.0006 & 0.0006 & 0.0012 & 0.0008 & 0.0008 & 0.0012 \\
pert. error     & 0.0035 & 0.0021 & 0.0024 & 0.0035 & 0.0033 & 0.0039 \\
hadr. error     & 0.0013 & 0.0006 & 0.0020 & 0.0010 & 0.0007 & 0.0010 \\
total error     & 0.0085 & 0.0100 & 0.0091 & 0.0074 & 0.0072 & 0.0081 \\\hline
fit range     & 0.75-0.94 & 1.6-4.4 & 0.08-0.25 & 0.22-0.75 & 0.08-0.25 & 0.11-0.35 \\ \hline
\end{tabular}
}

\renewcommand{\arraystretch}{1.1}
\renewcommand{\textfraction}{0.0001}
\TABLE{
\caption[Individual results] {\label{tab:indiv2}{\small Results for
$\alpha_s(Q)$ as obtained from NNLO+NLLA fits to distributions of
event-shape variables at $Q = \sqrt{s} = $183, 189, 200 and 206 GeV.}}
\begin{tabular}{|l|c|c|c|c|c|c|}\hline
\multicolumn{7}{|c|}{ $Q$ = 183 GeV} \\ \hline
variable          & $T$   & $-\ln y_3$ & $M_H$ & $C$ & $B_W$ & $B_T$  \\ \hline
$\alpha_s$        & 0.1131 & 0.1083 & 0.1129 & 0.1094 & 0.1091 & 0.1148 \\
stat. error       & 0.0036 & 0.0050 & 0.0038 & 0.0032 & 0.0027 & 0.0030 \\
exp. error        & 0.0007 & 0.0007 & 0.0012 & 0.0011 & 0.0008 & 0.0011 \\
pert. error       & 0.0034 & 0.0021 & 0.0023 & 0.0034 & 0.0033 & 0.0037 \\
hadr. error       & 0.0013 & 0.0005 & 0.0018 & 0.0010 & 0.0007 & 0.0010 \\
total error       & 0.0051 & 0.0055 & 0.0050 & 0.0049 & 0.0043 & 0.0050 \\ \hline
fit range     & 0.80-0.96 & 2.4-4.8 & 0.06-0.20 & 0.22-0.60 & 0.065-0.20 & 0.11-0.30 \\ \hline
\multicolumn{7}{|c|}{$Q$ = 189 GeV} \\ \hline
variable          &  $T$   & $-\ln y_3$ & $M_H$ & $C$ & $B_W$ & $B_T$  \\ \hline
$\alpha_s$        & 0.1119 & 0.1087 & 0.1087 & 0.1121 & 0.1056 & 0.1137 \\
stat. error       & 0.0026 & 0.0031 & 0.0032 & 0.0018 & 0.0019 & 0.0019 \\
exp. error        & 0.0007 & 0.0005 & 0.0017 & 0.0009 & 0.0009 & 0.0012 \\
pert. error       & 0.0033 & 0.0020 & 0.0023 & 0.0034 & 0.0032 & 0.0037 \\
hadr. error       & 0.0012 & 0.0005 & 0.0018 & 0.0010 & 0.0006 & 0.0010 \\
total error       & 0.0045 & 0.0038 & 0.0046 & 0.0040 & 0.0039 & 0.0044 \\ \hline
fit range     & 0.80-0.96 & 2.4-4.8 & 0.06-0.20 & 0.22-0.60 & 0.065-0.20 & 0.11-0.30 \\ \hline
\multicolumn{7}{|c|}{$Q$ = 200 GeV} \\ \hline
variable          &  $T$   & $-\ln y_3$ & $M_H$ & $C$ & $B_W$ & $B_T$  \\ \hline
$\alpha_s$        & 0.1078 & 0.1065 & 0.1020 & 0.1109 & 0.1047 & 0.1081 \\
stat. error       & 0.0027 & 0.0032 & 0.0034 & 0.0021 & 0.0019 & 0.0024 \\
exp. error        & 0.0007 & 0.0005 & 0.0019 & 0.0008 & 0.0008 & 0.0013 \\
pert. error       & 0.0032 & 0.0020 & 0.0023 & 0.0033 & 0.0032 & 0.0036 \\
hadr. error       & 0.0012 & 0.0005 & 0.0016 & 0.0009 & 0.0006 & 0.0010 \\
total error       & 0.0045 & 0.0038 & 0.0048 & 0.0041 & 0.0039 & 0.0046 \\ \hline
fit range     & 0.80-0.96 & 2.4-4.8 & 0.06-0.20 & 0.22-0.60 & 0.065-0.20 & 0.11-0.30 \\ \hline
\multicolumn{7}{|c|}{$Q$ = 206 GeV} \\ \hline
variable          &  $T$   & $-\ln y_3$ & $M_H$ & $C$ & $B_W$ & $B_T$  \\ \hline
$\alpha_s$        & 0.1084 & 0.1040 & 0.1076 & 0.1076 & 0.1051 & 0.1089 \\
stat. error       & 0.0025 & 0.0032 & 0.0024 & 0.0019 & 0.0017 & 0.0020 \\
exp. error        & 0.0007 & 0.0005 & 0.0012 & 0.0008 & 0.0008 & 0.0011 \\
pert. error       & 0.0032 & 0.0020 & 0.0022 & 0.0033 & 0.0031 & 0.0035 \\
hadr. error       & 0.0012 & 0.0004 & 0.0016 & 0.0009 & 0.0006 & 0.0010 \\
total error       & 0.0043 & 0.0038 & 0.0039 & 0.0040 & 0.0037 & 0.0043 \\ \hline
fit range     & 0.80-0.96 & 2.4-4.8 & 0.04-0.20 & 0.22-0.60 & 0.05-0.20 & 0.11-0.30 \\ \hline
\end{tabular}
}

\TABLE{
\caption[Some fits] {\label{tab:comp_nlo_nnlo}{\small Fit results for $\asmz$ using
different predictions of perturbative QCD, with the renormalisation
scale fixed to $\mu = \Mz$.}}
\begin{tabular}{|l|c|c|c|c|c|c|}\hline
          & $T$         & $C$         & $M_H$       & $B_W$       & $B_T$ & $-\ln y_3$\\ \hline
NNLO+NLLA & 0.1266      & 0.1252      & 0.1211      & 0.1196      & 0.1268 & 0.1186 \\
$\chi^2/N_{dof}$ & 0.16 & 0.47        & 4.4         & 4.4         & 0.84   & 1.89   \\
stat.error & 0.0002     & 0.0002      & 0.0003      & 0.0002      & 0.0002 & 0.0002 \\ \hline
NLO+NLLA  & 0.1282      & 0.1244      & 0.1180      & 0.1161      & 0.1290 & 0.1187 \\
$\chi^2/N_{dof}$ & 0.74 & 1.88        & 14.5        & 19.6        & 9.7    & 4.7   \\
stat.error & 0.0002     & 0.0002      & 0.0003      & 0.0002      & 0.0002 & 0.0002\\ \hline
NNLO      & 0.1275      & 0.1273      & 0.1248      & 0.1242      & 0.1279 & 0.1192 \\
$\chi^2/N_{dof}$ & 1.16 & 1.08        & 4.1         & 2.74        & 0.50   & 1.17  \\
stat.error & 0.0002     & 0.0002      & 0.0004      & 0.0002      & 0.0002 & 0.0003\\ \hline
fit range & 0.75 - 0.91 & 0.36 - 0.74 & 0.10 - 0.22 & 0.09 - 0.19 & 0.16 - 0.30 & 1.6-4.0 \\ \hline
\end{tabular}
}

\TABLE{
\caption[Individual results] {\label{tab:Combi}{\small Combined results for
$\alpha_s(Q)$ using NNLO+NLLA predictions.}}
\begin{tabular}{|l|cccccccc|}\hline
$Q\; [$GeV$]$    & 91.2   & 133    & 161    & 172    & 183    & 189    & 200    & 206 \\ \hline
$\alpha_s(Q)$  & 0.1221 & 0.1179 & 0.1201 & 0.1086 & 0.1112 & 0.1099 & 0.1067 & 0.1066 \\
stat. error    & 0.0001 & 0.0029 & 0.0043 & 0.0052 & 0.0023 & 0.0016 & 0.0017 & 0.0015 \\
exp. error     & 0.0008 & 0.0008 & 0.0008 & 0.0008 & 0.0009 & 0.0009 & 0.0008 & 0.0009 \\
pert. error    & 0.0041 & 0.0036 & 0.0033 & 0.0032 & 0.0031 & 0.0030 & 0.0029 & 0.0029 \\
hadr. error    & 0.0018 & 0.0012 & 0.0010 & 0.0010 & 0.0009 & 0.0008 & 0.0008 & 0.0008 \\
total error    & 0.0045 & 0.0049 & 0.0056 & 0.0062 & 0.0040 & 0.0036 & 0.0036 & 0.0034 \\ \hline
RMS            & 0.0038 & 0.0023 & 0.0026 & 0.0029 & 0.0027 & 0.0030 & 0.0030 & 0.0019 \\ \hline
\end{tabular}
}
\TABLE{
\caption[Individual results] {\label{tab:combz}{\small Combined results for
$\asmz$ using NNLO+NLLA predictions.}}
\begin{tabular}{|l|cccccccc|}\hline
$Q\; [$GeV$]$   & 91.2   & 133    & 161    & 172    & 183    & 189    & 200    & 206 \\ \hline
$\asmz$         & 0.1221 & 0.1251 & 0.1316 & 0.1190 & 0.1235 & 0.1225 & 0.1196 & 0.1200 \\
stat. error     & 0.0001 & 0.0033 & 0.0052 & 0.0063 & 0.0028 & 0.0020 & 0.0022 & 0.0019 \\
exp. error      & 0.0008 & 0.0010 & 0.0010 & 0.0011 & 0.0011 & 0.0011 & 0.0011 & 0.0011 \\
pert. error     & 0.0041 & 0.0038 & 0.0036 & 0.0035 & 0.0034 & 0.0033 & 0.0033 & 0.0032 \\
hadr. error     & 0.0018 & 0.0014 & 0.0012 & 0.0011 & 0.0011 & 0.0010 & 0.0010 & 0.0010 \\
total error     & 0.0045 & 0.0053 & 0.0065 & 0.0074 & 0.0047 & 0.0042 & 0.0042 & 0.0041 \\ \hline
RMS             & 0.0038 & 0.0026 & 0.0032 & 0.0035 & 0.0033 & 0.0037 & 0.0038 & 0.0024 \\ \hline
\end{tabular}
}

\TABLE{
\caption[Individual results] {\label{tab:combz_nnlo}{\small Combined results for
$\asmz$ using NNLO predictions.}}
\begin{tabular}{|l|cccccccc|}\hline
$Q\; [$GeV$]$   & 91.2   & 133    & 161    & 172    & 183    & 189    & 200    & 206 \\ \hline
$\asmz$         & 0.1239 & 0.1270 & 0.1313 & 0.1192 & 0.1226 & 0.1234 & 0.1200 & 0.1202 \\
stat. error     & 0.0002 & 0.0033 & 0.0051 & 0.0063 & 0.0028 & 0.0020 & 0.0021 & 0.0019 \\
exp. error      & 0.0009 & 0.0009 & 0.0009 & 0.0010 & 0.0010 & 0.0010 & 0.0010 & 0.0010 \\
pert. error     & 0.0030 & 0.0030 & 0.0028 & 0.0028 & 0.0027 & 0.0026 & 0.0025 & 0.0025 \\
hadr. error     & 0.0018 & 0.0014 & 0.0012 & 0.0012 & 0.0011 & 0.0010 & 0.0010 & 0.0010 \\
total error     & 0.0037 & 0.0048 & 0.0060 & 0.0070 & 0.0041 & 0.0036 & 0.0036 & 0.0034 \\ \hline
RMS             & 0.0036 & 0.0014 & 0.0043 & 0.0019 & 0.0027 & 0.0027 & 0.0034 & 0.0024 \\ \hline
\end{tabular}
}

\TABLE[h]{
\caption[Individual results] {\label{tab:comblep}{\small Weighted average of
combined measurements for $\asmz$ obtained at energies from 91.2 GeV
to 206 GeV and the average without the point at $\sqrt{s} = \Mz$ using NNLO+NLLA predictions.}}
\hspace{1cm}
\begin{tabular}{|l|cc|}\hline
data set        & LEP1 + LEP2  & LEP2 \\ \hline
$\asmz$         & 0.1224       & 0.1224 \\
stat. error     & 0.0009       & 0.0011 \\
exp. error      & 0.0009       & 0.0010 \\
pert. error     & 0.0035       & 0.0034 \\
hadr. error     & 0.0012       & 0.0011 \\
total error     & 0.0039       & 0.0039 \\ \hline
\end{tabular}
\hspace{1cm}
}

\renewcommand{\arraystretch}{1.1}
\renewcommand{\textfraction}{0.0001}
\TABLE{
\caption[published results] {\label{tab:compall}{\small Comparison of
combined results obtained with different theoretical predictions on $\asmz$ using ALEPH data at energies from 91.2 GeV
to 206 GeV.}}
\hspace{1cm}
\begin{tabular}{|l|ccc|}\hline
theory input    & NNLO+NLLA    & NNLO & NLO+NLLA \\ \hline
$\asmz$         & 0.1224       & 0.1228 & 0.1215\\
stat. error     & 0.0009       & 0.0008 & 0.0010 \\
exp. error      & 0.0009       & 0.0009 & 0.0009 \\
pert. error     & 0.0035       & 0.0027 & 0.0053 \\
hadr. error     & 0.0012       & 0.0012 & 0.0012 \\
total error     & 0.0039       & 0.0032 & 0.0056 \\ \hline
\end{tabular}
\hspace{1cm}
}

\renewcommand{\arraystretch}{1.1}
\renewcommand{\textfraction}{0.0001}
\TABLE{
\caption[LogRmu]{\label{tab:logrmu}{\small
Comparison of the theoretical systematic uncertainties for the $\ln R$ and $\ln R(\mu)$ matching schemes. 
Only the uncertainty for missing higher orders as  
obtained from the uncertainty band method are included, using $\asmz$=0.1224. 
The total perturbative 
uncertainty also accounts for the mass corrections, the latter are the same for both matching schemes.}}
\begin{tabular}{|l|c|c|c|c|c|c|}\hline
variable           & $T$  & $-\ln y_3$ & $M_H$& $C$    & $B_W$  & $B_T$  \\ \hline
$\ln R(\mu)$       & 0.0017 & 0.0028 & 0.0025 & 0.0030 & 0.0031 & 0.0025 \\
$\ln R$            & 0.0047 & 0.0029 & 0.0033 & 0.0049 & 0.0045 & 0.0053 \\ \hline
\end{tabular}
}

\clearpage

\TABLE{
\caption[Individual results] {\label{tab:comblogrmu}{\small Weighted average of
combined measurements for $\asmz$, obtained at energies from 91.2 GeV
to 206 GeV and without the point at $\sqrt{s} = \Mz$, using in all cases the
$\ln R(\mu)$ matching scheme.}}
\hspace{1cm}
\begin{tabular}{|l|cc|}\hline
data set        & LEP1 + LEP2  & LEP2 \\ \hline
$\asmz$         & 0.1227       & 0.1226 \\
stat. error     & 0.0008       & 0.0010 \\
exp. error      & 0.0009       & 0.0010 \\
pert. error     & 0.0022       & 0.0021 \\
hadr. error     & 0.0012       & 0.0011 \\
total error     & 0.0028       & 0.0028 \\ \hline
\end{tabular}
\hspace{1cm}
}

\renewcommand{\arraystretch}{1.1}
\renewcommand{\textfraction}{0.0001}
\TABLE{
\caption[norm]{\label{tab:norm}{\small
Results on $\asmz$ from LEP1 data using different normalisation and mass correction schemes.}}
\begin{tabular}{|l|c|c|c|c|c|c|}\hline
variable               & $T$   & $-\ln y_3$ & $M_H$ & $C$ & $B_W$ & $B_T$  \\ \hline
$\sigma_{\mathrm{had}}$ (massive)    & 0.1266 & 0.1186 & 0.1211 & 0.1252 & 0.1196 & 0.1268 \\
$\sigma_{\mathrm{had}}$ (massless)   & 0.1266 & 0.1187 & 0.1212 & 0.1252 & 0.1196 & 0.1268 \\
massless expansion, massive $A, B$   & 0.1260 & 0.1183 & 0.1208 & 0.1247 & 0.1192 & 0.1262 \\
massless expansion, massless $A, B$  & 0.1256 & 0.1179 & 0.1215 & 0.1242 & 0.1188 & 0.1253 \\ 
$\sigma_{\mathrm{had}}$ (massive, $M_{\rm b}$ = 4.0\, \GeVcc\ )    & 0.1264 & 0.1185 & 0.1212 & 0.1251 & 0.1195 & 0.1267 \\
$\sigma_{\mathrm{had}}$ (massive, $M_{\rm b}$ = 5.0\, \GeVcc\ )    & 0.1268 & 0.1189 & 0.1210 & 0.1252 & 0.1198 & 0.1270 \\ \hline
\end{tabular}
}

\TABLE{
\caption[Individual results] {\label{tab:weights}{\small Weights (in per cent) of the different
centre-of-mass energy points in the global combination, with and without the inclusion of theoretical uncertainties.}}
\begin{tabular}{|l|cccccccc|}\hline
$Q\; [$GeV$]$    & 91.2 & 133 & 161 & 172 & 183 & 189 & 200 & 206 \\ \hline
with pert.err. & 14.0 & 10.3 & 6.8 & 5.3 & 13.1 & 16.7 & 16.4 & 17.5 \\
w/o pert.err.  & 80.0 & 2.5 & 2.6 & 2.5 & 2.8 & 3.1 & 3.2 & 3.3 \\ \hline
\end{tabular}
}

\TABLE{
\caption[Individual results] {\label{tab:combwgt}{\small Weighted average of the
combined measurements for $\asmz$, based on weights which do not include the
theoretical uncertainty.}}
\hspace{1cm}
\begin{tabular}{|l|cc|}\hline
data set        & LEP1 + LEP2  & LEP2 \\ \hline
$\asmz$         & 0.1222       & 0.1228 \\
stat. error     & 0.0003       & 0.0013 \\
exp. error      & 0.0007       & 0.0010 \\
pert. error     & 0.0039       & 0.0034 \\
hadr. error     & 0.0017       & 0.0011 \\
total error     & 0.0044       & 0.0040 \\ \hline
\end{tabular}
\hspace{1cm}
}

\TABLE{
\caption[Some fits] {\label{tab:hw++tune}{\small Comparison of hadron level predictions 
from various event generators to the ALEPH event-shape data.
The Table shows the $\chi^2$ values normalised 
to the number of experimental bins, including statistical and experimental systematic 
uncertainties of the data.}}
\begin{tabular}{|l|c|c|c|c|c|c|c|}\hline
            & $T$         & $C$         & $M_H$       & $B_W$       & $B_T$ & $-\ln y_3$ & global \\ \hline
PYTHIA      & 0.55      & 1.05      & 1.9       & 2.5       & 0.57 & 1.31 & 1.30\\
ARIADNE     & 0.44      & 0.75      & 0.60      & 0.97      & 0.58 & 0.52 & 0.65 \\
HERWIG      & 6.9       & 5.3       & 9.4       & 9.8       & 4.4  & 4.0  & 6.6 \\
HW++        & 17.5      & 16.5      & 18.2      & 13.4      & 9.4  & 5.6  & 13.5\\
HW++ MCNLO  & 9.6       & 15.9      & 9.2       & 10.5      & 11.8 & 5.5  & 10.4\\
HW++ POWHEG & 3.9       & 11.2      & 8.5       & 6.9       & 3.5  & 2.3  & 6.1 \\ \hline
\end{tabular}
}

\TABLE{
\caption[Some fits] {\label{tab:hw++fits}{\small Fit results for $\asmz$ using LEP1 data and NLLO+NLLA but 
different hadronisation corrections. In all cases the same detector corrections, obtained from a full detector simulation 
using PYTHIA as generator is applied. The statistical errors are essentially unaltered compared to those in Table \ref{tab:comp_nlo_nnlo}.}}
\begin{tabular}{|l|c|c|c|c|c|c|}\hline
   $\asmz$       & $T$         & $C$         & $M_H$       & $B_W$       & $B_T$ & $-\ln y_3$\\ \hline
PYTHIA  & 0.1266       & 0.1252      & 0.1211      & 0.1196      & 0.1268 & 0.1186 \\
$\chi^2/N_{dof}$ & 0.16 & 0.47        & 4.4         & 4.4         & 0.84   & 1.89   \\ \hline
ARIADNE   & 0.1285     & 0.1268      & 0.1234      & 0.1212      & 0.1258 & 0.1202 \\
$\chi^2/N_{dof}$ & 0.96 & 0.52        & 2.5        & 3.1         & 2.15   & 1.41   \\ \hline
HERWIG    & 0.1256      & 0.1242     & 0.1253      & 0.1203      & 0.1258 & 0.1203 \\
$\chi^2/N_{dof}$ & 0.5 & 0.65        & 4.4         & 2.0        & 2.15   & 0.8  \\ \hline
HW++      & 0.1242      & 0.1228     & 0.1299      & 0.1212      & 0.1238 & 0.1168 \\
$\chi^2/N_{dof}$ & 6.6 & 3.2        & 3.3         & 1.33        & 2.65   & 0.56  \\ \hline
HW++ MCNLO      & 0.1234      & 0.1220     & 0.1292      & 0.1220      & 0.1232 & 0.1175 \\
$\chi^2/N_{dof}$ & 10.7 & 4.2        & 2.2         & 1.1        & 5.7   & 0.69  \\ \hline
HW++ POWHEG      & 0.1189      & 0.1179     & 0.1236      & 0.1169      & 0.1224 & 0.1142 \\
$\chi^2/N_{dof}$ & 1.46 & 2.55        & 3.8         & 3.9        & 1.54   & 0.56  \\ \hline
\end{tabular}
}

\end{document}